\documentclass[tradiabstract]{aa}

\newcommand{\bc}{\begin{center}}
\newcommand{\ec}{\end{center}}
\newcommand{\ben}{\begin{enumerate}}
\newcommand{\een}{\end{enumerate}}
\newcommand{\bd}{\begin{description}}
\newcommand{\ed}{\end{description}}
\newcommand{\bmi}[1]{\begin{minipage}{#1 cm}}
\newcommand{\emi}{\end{minipage}}
\newcommand{\bmif}[1]{\begin{minipage}{#1\textwidth}}

\def\llabel#1{\label{sc:#1}}
\def\elabel#1{\label{eq:#1}}

\def\eck#1{\left\lbrack #1 \right\rbrack}

\def\rund#1{\left( #1 \right)}
\def\abs#1{\left\vert #1 \right\vert}

\def\ave#1{\left\langle #1 \right\rangle}

\def\U{{\cal U}}
\def\d{{\rm d}}

\def\eps{{\epsilon}}

\def\vp{\varphi}
\def\vt{{\vartheta}}

\def\Real{{\rm I\mathchoice{\kern-0.70mm}{\kern-0.70mm}{\kern-0.65mm}%
  {\kern-0.50mm}R}}
\def\C{\rm C\kern-.42em\vrule width.03em height.58em depth-.02em
       \kern.4em}

\def\bx#1{\leavevmode\thinspace\hbox{\vrule\vtop{\vbox{\hrule\kern1pt
        \hbox{\vphantom{\tt/}\thinspace{\bf#1}\thinspace}}
      \kern1pt\hrule}\vrule}\thinspace}

\def\vc#1{{\mbox{\boldmath$#1$\unboldmath}}}
{\catcode`\@=11
\gdef\SchlangeUnter#1#2{\lower2pt\vbox{\baselineskip 0pt \lineskip0pt
  \ialign{$\m@th#1\hfil##\hfil$\crcr#2\crcr\sim\crcr}}}
}

\def\ueber#1#2{{\setbox0=\hbox{$#1$}%
  \setbox1=\hbox to\wd0{\hss$\scriptscriptstyle #2$\hss}%
  \offinterlineskip
  \vbox{\box1\kern0.4mm\box0}}{}}

\def\bx#1{\leavevmode\thinspace\hbox{\vrule\vtop{\vbox{\hrule\kern1pt
        \hbox{\vphantom{\tt/}\thinspace{\bf#1}\thinspace}}
      \kern1pt\hrule}\vrule}\thinspace}

{\catcode`\@=11
\gdef\SchlangeUnter#1#2{\lower2pt\vbox{\baselineskip 0pt \lineskip0pt
  \ialign{$\m@th#1\hfil##\hfil$\crcr#2\crcr\sim\crcr}}}
}

\def\ts{\thinspace}

\newcommand{\tvec}{\ensuremath{\vc{\theta}}}
\newcommand{\tE}{\ensuremath{\theta_\mathrm E}}

\newcommand{\tc}{\ensuremath{\theta_\mathrm c}}

\newcommand{\avec}{\ensuremath{\vc{\alpha}}}
\newcommand{\ahat}{\ensuremath{\vc{\hat{\alpha}}}}
\newcommand{\atilde}{\ensuremath{\vc{\tilde{\alpha}}}}

\newcommand{\bvec}{\ensuremath{\vc{\beta}}}
\newcommand{\bhat}{\ensuremath{\vc{\hat{\beta}}}}
\newcommand{\kbar}{\ensuremath{\bar{\kappa}}}
\newcommand{\khat}{\ensuremath{\hat{\kappa}}}
\newcommand{\ki}{\ensuremath{\hat{\kappa}_\mathrm{I}}}
\newcommand{\ktilde}{\ensuremath{\tilde{\kappa}}}
\newcommand{\phat}{\ensuremath{\hat{\psi}}}
\newcommand{\ptilde}{\ensuremath{\tilde{\psi}}}
\newcommand{\Ahat}{\ensuremath{\hat{\mathcal{A}}}}
\newcommand{\gp}{\ensuremath{\mathcal{\gamma_\mathrm{p}}}}
\newcommand{\gm}{\ensuremath{\mathcal{\gamma_\mathrm{m}}}}
\newcommand{\ealpha}{\ensuremath{\Delta \avec}}
\newcommand{\eacc}{\ensuremath{\varepsilon_\mathrm{acc}}}

\newcommand{\N}{\ensuremath{\mathcal{N}}}

\usepackage{amsmath}
\usepackage[english=british]{csquotes}

\usepackage{graphicx}
\usepackage{txfonts}
\usepackage{natbib}
	\bibpunct{(}{)}{;}{a}{}{,}
\usepackage{subfigure}
\usepackage{enumerate}
\begin{document} 
\title{Ambiguities in gravitational lens models: the density field
  from the source position transformation }
\author{Sandra Unruh\inst{1} \and Peter Schneider \inst{1} \and Dominique Sluse
          \inst{2}}
\institute{Argelander-Institut f\"ur Astronomie, Universit\"at
	Bonn, Auf dem H\"ugel 71, D-53121 Bonn, Germany\\
	sandra, peter@astro.uni-bonn.de
\and
	STAR Institute, Quartier Agora, All\'ee du six Ao\^ut, 19c, University of Li\`ege, B-4000 Li\`ege, Belgium\\
	dsluse@ulg.ac.be}
%
%
%
%
%
  \abstract
  {Strong gravitational lensing is regarded as the most precise
  technique to measure the mass in the inner region of galaxies or
  galaxy clusters. In particular, the mass within one Einstein radius
  can be determined with an accuracy of order of a few percent or
  better, depending on the image configuration. For other radii,
  however, degeneracies exist between galaxy density profiles,
  precluding an accurate determination of the enclosed mass.  The
  source position transformation (SPT), which includes the well-known
  mass-sheet transformation (MST) as a special case, describes this
  degeneracy of the lensing observables in a more general way. In this
  paper we explore properties of an SPT, removing the MST to leading
  order, i.e., we consider degeneracies which have not been described
  before.  The deflection field $\ahat(\vc\theta)$ resulting from an
  SPT is not curl-free in general, and thus not a deflection that can
  be obtained from a lensing mass distribution. Starting from a
  variational principle, we construct lensing potentials that give
  rise to a deflection field $\atilde$, which differs from $\ahat$ by
  less than an observationally motivated upper limit. The corresponding
  mass distributions from these `valid' SPTs are studied: their radial
  profiles are modified relative to the original mass distribution in
  a significant and non-trivial way, and originally axi-symmetric mass
  distributions can obtain a finite ellipticity. These results
  indicate a significant effect of the SPT on quantitative analyses of
  lens systems. We show that the mass inside the Einstein radius of
  the original mass distribution is conserved by the SPT; hence, as is
  the case for the MST, the SPT does not affect the mass determination
  at the Einstein radius. Furthermore, we analyse a degeneracy between
  two lens models, empirically found previously, and show that this
  degeneracy can be interpreted as being due to an SPT. Thus,
  degeneracies between lensing mass distributions are not just a
  theoretical possibility, but do arise in actual lens modeling.}

\titlerunning{Source-position transformation} 

\maketitle

%
%
\section{\llabel{Sc1}Introduction}
Strong gravitational lensing provides a highly valuable tool to obtain mass properties of galaxies and galaxy clusters \citep[see, e.g.][and references therein]{Bart10,SaasFee3}. In particular, multiple image systems yield strong constraints on the mass distribution. The mass enclosed within the Einstein radius presents the most robust galaxy mass estimate currently available. Furthermore, the shape of the mass distribution (e.g. ellipticity, orientation) is well defined.

However, mass estimates for radii smaller or larger than the Einstein radius are less accurate. If only a finite set of individual lensed compact images is observed, too few observational constraints are available and certainly no unique radial mass profile can be found. The situation changes somewhat if extended source components are lensed where the constraints on the mass distribution are much more stringent. Nonetheless, even if we could find a mass model which reproduces all constraints perfectly, such a mass model would not be unique either. The reason for this degeneracy is known since 1985 \citep{FGS85} and is called the mass-sheet transformation (MST). If a given surface mass density \(\kappa (\tvec)\) reproduces all observational constraints, then the whole family of mass models,
\begin{equation}
	\kappa_\lambda(\vc\theta)=\lambda\kappa(\vc\theta)+(1-\lambda)\;,
	\elabel{MST}
\end{equation}
will do the same. In particular, the MST leaves all observables invariant except the time delay\footnote{although time delay ratios stay constant.}. The transformation (\ref{eq:MST}) modifies the slope of the density profile with a constant factor \(\lambda\). This affects mass measurements outside the Einstein radius \tE \ and determination of the Hubble constant \(H_0\) directly.

\citet[][hereafter SS13]{SS13} presented two mass profiles (namely, a Hernquist profile
plus a modified Navarro, Frank and White profile, as well as a
power-law mass profile) which showed almost the same imaging
properties, although they are not exactly related through an
MST. Following this unexpected result it became apparent that an even
more general invariance transformation than the MST exists. The
so-called source-position transformation (SPT) was finally introduced
in \citet[][hereafter SS14]{SS14}.

For isolated individual images many ambiguities for the lens equation
exist. Local transformations of the lensing mass distribution, which
still reproduce the positional constraints from the lensed images,
lead to an infinite number of mass models \citep[see
  e.g.][]{Saha97,Diego05,Coe08,Liesenborgs2012}. The MST as given in
Eq.\ts (\ref{eq:MST}) is a {\it global} transformation and equivalent
to an isotropic uniform stretching of the source plane by a constant
factor \(\lambda\). The SPT is based on a more general (global)
transformation of the source plane coordinates. Such transformations
\(\bhat (\bvec)\), where \bhat \ denotes the transformed source
position, give rise to a new deflection law \(\ahat (\tvec) = \tvec -
\bhat ( \tvec - \avec(\tvec) )\). The new deflection law \ahat \ will
in general not be a gradient field and thus cannot be 
obtained from the deflection caused by a lens. However, if the curl
component of $\ahat$ is
sufficiently small, then one may find a lensing mass distribution
which yields a deflection law which is very close to $\ahat$, so close
that it cannot be observationally distinguished from $\ahat$. In this
paper we will explore this possibility, which of course depends on
the SPT $\bhat(\bvec)$. In particular, if this deformation is `too
strong', then the resulting $\ahat$ cannot be approximated with the
deflection due to a lens -- this will restrict the freedom in choosing
transformations $\bhat(\bvec)$.

The outline of the paper is as follows. In Sect.\ts\ref{sc:Sc2} we
will recapitulate the principle of the SPT.  We characterize the
deviation of the deflection law from a gradient field quantitatively
in Sect.\ts\ref{sc:Sc3} by finding a gravitational potential \ptilde \
such that \(\atilde = \nabla \ptilde\) is as close as possible to the
SPT-transformed deflection law \ahat. To do so, we will start from a
variational principle and show that the modified deflection potential
\ptilde \ has to fulfill von Neumann boundary conditions. Those can be
solved using a Green's function, and the solution will be given
explicitly for a circular region. Furthermore, a numerical
approach will be presented to find
degenerate deflection laws and their corresponding mass profiles. By
considering a specific deformation function \(\bhat (\bvec)\) and
assuming a positional accuracy on lensed image positions typical of
the Hubble Space Telescope (HST), we will present in
Sect.\ts\ref{sc:Sc4} the implications of the `allowed' SPTs on current
mass profile determinations, regarding the radial mass profile and the
angular structure of the lens.  Different diagnostics for the change
of the mass profile by an SPT, and how it can be distinguished from an
MST, will be explored in Sect.\ts\ref{sc:Sc5} in terms of the aperture
mass. Finally, we will discuss our findings in Sect.\ts\ref{sc:Sc6}.
\section{\llabel{Sc2}The principle of the source position
  transformation} 
In the following we will describe the principle of the SPT and its
properties. For a more detailed account the reader is referred to
SS14. We use standard gravitational lensing notation throughout this
paper \citep[see, e.g.,][]{SaasFee1}.

\begin{figure}[htbp]
	\centering
	\includegraphics[width=.45\textwidth]{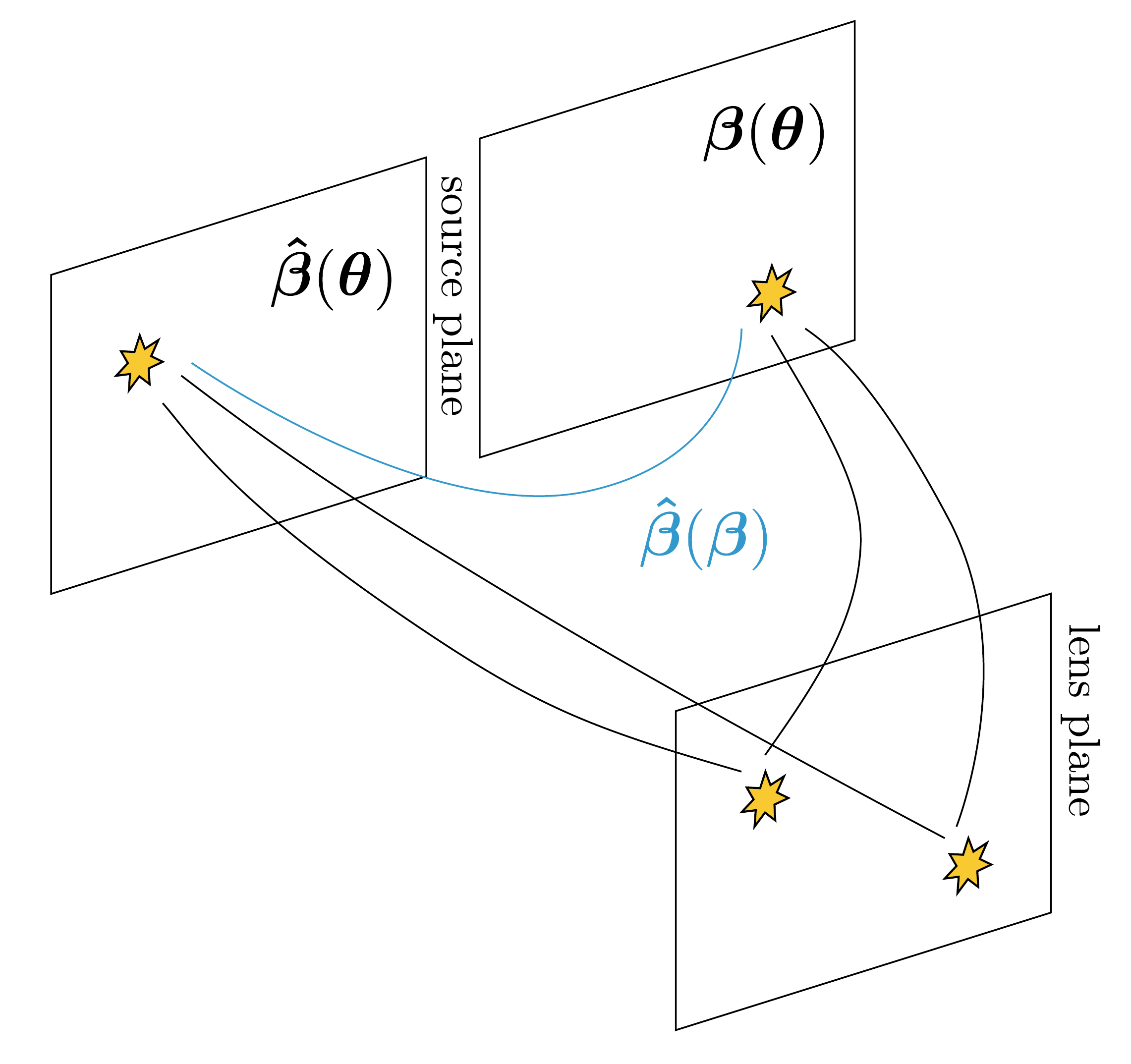}
	\caption{An illustration of the source position
          transformation. A source at $\bvec$ causes multiple images
          $\tvec$ in the lens plane under the deflection law
          $\avec$. The same multiple images are obtained from a
          source at $\bhat(\bvec)$, provided the deflection law is
          changed to $\ahat$, according to Eq.\ts(\ref{eq:hatalpha}).}
	\label{pic:SPT}
\end{figure}

In general, a surface mass density distribution \(\kappa (\tvec)\)
gives rise to a deflection law \(\avec(\tvec)\), where \tvec \ is the
angular position in the lens plane, i.e., the observer's sky. The mass
distribution or convergence \(\kappa\) is defined as the ratio of
projected surface mass density to the critical surface mass density, where
the latter depends only on the angular diameter distances of lens and
source. If that mass distribution is sufficiently concentrated (i.e.,
typically \(\kappa(\tvec) \gtrsim 1\) for some region in the lens
plane) a source may have multiple images, depending on its position
relative to the deflector on the sky. Then, the source located at the
(unobservable) position \bvec \ will have its images at locations
described by the solutions \(\tvec_i = \bvec + \avec (\tvec_i)\) of
the lens equation. Since multiple images are from the same source, we
can deduce the constraints on the deflection law \(\avec(\tvec)\) as
\begin{equation}
	\vc\theta_i-\vc\alpha(\vc\theta_i)=\vc\theta_j-\vc\alpha(\vc\theta_j)\;,
	\elabel{ta}
\end{equation}
or likewise for an alternative deflection law $\ahat(\tvec)$ 
as
%
\begin{equation}
	\vc\theta_i-\hat{\vc\alpha}(\vc\theta_i)
	=\vc\theta_j-\hat{\vc\alpha}(\vc\theta_j)\;,
	\elabel{tahat}
\end{equation}
for all $i<j$, such that \(\avec(\tvec)\) as well as \(\ahat(\tvec)\)
yield exactly the same sets of multiple images.  If such equivalent
deflection laws exist, they will correspond to source positions
\(\bvec = \tvec - \avec (\tvec)\) or \(\bhat = \tvec - \ahat(\tvec)\),
respectively (see Fig.~\ref{pic:SPT}).

We can now consider a one-to-one mapping \(\bhat(\bvec)\) that connects the original source coordinates to the new ones. This allows us to define the transformed deflection law as
\begin{equation}
	\hat{\vc\alpha}(\vc\theta)=\vc\alpha(\vc\theta)
	+\vc\beta-\hat{\vc\beta} (\bvec)
	=\vc\theta-\hat{\vc\beta}(\vc\theta-\vc\alpha(\vc\theta))\;,
	\elabel{hatalpha}
\end{equation}
where in the last step we inserted the original lens equation $\vc\beta=\vc\theta-\vc\alpha(\vc\theta)$.

Hence, any bijective (i.e., one-to-one) function \(\bhat(\bvec)\)
leads to an SPT which leaves the condition (\ref{eq:ta})
invariant. Moreover, as can be deduced from the Jacobian \(\Ahat
= \partial \bhat / \partial \tvec = (\partial \bhat / \partial \bvec)
(\partial \bvec / \partial \tvec) \) of the modified lens equation,
the relative magnification matrices and the relative image shapes
between image pairs of the same source \bhat \ remain
unchanged. However, the Jacobian \Ahat \ will not be symmetric in
general, and therefore \ahat \ cannot be written as the gradient of a
deflection potential \phat \ (i.e., \ahat \ is not a curl-free
field). This implies that no corresponding mass distribution \khat \
exists that yields a deflection angle \ahat, in general. However, it
was shown in SS14 that the asymmetric part of the Jacobian can be
small in realistic cases; this will be explored more quantitatively in
Sect.\ts\ref{sc:Sc3}. In the special case that the lens is
axisymmetric and the transformation $\bhat(\bvec)$ corresponds to a
radial stretching of the form
\begin{equation}
\bhat=f(|\bvec|)\bvec\;,
\elabel{radstretch}
\end{equation}
the SPT {\it is} an exact invariance transformation: in this case, the
Jacobian $\Ahat$ is symmetric, and for every transformation
(\ref{eq:radstretch}) and its corresponding deflection law \ahat \
there exists a corresponding axi-symmetric mass distribution \khat.

Provided the curl component of \ahat \ is small, then we expect that
there exists a mass distribution $\tilde\kappa$ whose corresponding
deflection law $\tilde{\vc\alpha}$ will be very similar to $\ahat$, in
the sense that their difference is smaller than the astrometric
accuracy of current observations. In this case, the SPT will be, for
all practical purposes, a global invariance transformation for lenses.
\section{\llabel{Sc3}The transformed mass distribution}
\subsection{The general method}
Since the deflection law $\hat{\vc\alpha}$ (\ref{eq:hatalpha}) is not a gradient field, it does not
correspond to a deflection field caused by a gravitational
lens. However, if the curl component of $\hat{\vc\alpha}$ is
sufficiently small, one may be able to find a deflection potential
$\tilde \psi$ and a corresponding deflection law
$\tilde{\vc\alpha}=\nabla \tilde\psi$ such that the difference between
$\hat{\vc\alpha}$ and $\tilde{\vc\alpha}$ is small, e.g., smaller
than the astrometric accuracy of current observations. Since only the
region of the lens plane where multiple images occur is constrained by
lensing observations, the difference $\hat{\vc\alpha} -
\tilde{\vc\alpha}$ needs to be small only in a finite region, which we
denote as $\cal U$.

We thus consider the `action'
\begin{equation}
	S=\int_{\cal U} \d^2\theta\;\abs{\nabla \tilde\psi-\hat{\vc\alpha}}^2\;,
\end{equation}
for which we want to find a minimum. This is achieved by considering
small variations of $\tilde\psi \to \tilde\psi +\delta\tilde\psi$, and
finding the conditions for which the action is stationary for all
variations $\delta\tilde\psi$. Up to linear terms in
$\delta\tilde\psi$, we find 
\begin{eqnarray}
	S+\delta S&=&\int_{\cal U} \d^2\theta\;\abs{\nabla \tilde\psi +\nabla(\delta\tilde\psi)-\hat{\vc\alpha}}^2 \nonumber \\
	&=&S+2 \int_{\cal U} \d^2\theta\;\nabla(\delta\tilde\psi)
	\rund{\nabla \tilde\psi -\hat{\vc\alpha}} \nonumber \\
	&=&S+2\int_{\partial \cal U}\d s\;\delta\tilde\psi
	\rund{\nabla \tilde\psi -\hat{\vc\alpha}}\cdot \vc n \\
	&&-2\int_{\cal U} \d^2\theta\;\delta\tilde\psi
	\rund{\nabla^2\tilde\psi-\nabla\cdot\hat{\vc\alpha}}\;, \nonumber
\end{eqnarray}
where we made use of Gau\ss \ divergence theorem. The boundary curve
of $\cal U$ is denoted as $\partial \cal U$, $\d s$ is the line
element of the boundary curve, and $\vc n(s)$ the outward directed
normal vector. Requiring $\delta S=0$ leads to the von Neumann problem 
\begin{equation}
	\nabla^2\tilde\psi=\nabla\cdot\hat{\vc\alpha}=:2\hat\kappa \;; \;
	\hbox{and}\quad \nabla \tilde\psi \cdot \vc n = \hat{\vc\alpha}\cdot \vc n\;,
	\elabel{vNP}
\end{equation}
where the first equation is required for all points $\vc\theta\in
{\cal U}$, and the second one for all points on the boundary $\partial
\cal U$. The solution $\tilde\psi$ of Eq.\ts (\ref{eq:vNP}) is specified only
up to an additive constant, since a constant in the deflection
potential does not affect the deflection angle. 

In order to solve the system (\ref{eq:vNP}), we can either use
numerical standard methods for such boundary problems, or we can
obtain the solution by means of a Green's function. Both methods will
be explored in this section.
\subsection{Solving the von Neumann problem numerically}
\label{sec:solv_neumann}
We defined the convergence of the transformed deflection law to be
\(\khat = \nabla \cdot \ahat /2\). The curl component of $\ahat$ is
reasonably small if the closest curl-free approximation to \ahat \
(which is \atilde) is smaller than a chosen astrometric accuracy \eacc
\begin{equation}
	| \, \ahat (\tvec) - \atilde (\tvec) \, |  = | \ealpha (\tvec) | < \eacc
	\elabel{prec}
\end{equation}
for all $\tvec\in\cal{U}$.  To solve the system (\ref{eq:vNP})
numerically, we set up a successive overrelaxation method
\citep[SOR;][their Sect.~19.5]{Press96} on a square grid to calculate
\ptilde. An SOR is a converging iterative process based on the
extrapolation of the Gau\ss-Seidel method, and it is a standard method
to solve boundary value problems \citep[see, e.g.,][]{Seitz01}. Using
a second-order accurate finite differencing scheme, the deflection law
\atilde \ is then derived from the deflection potential \ptilde.

The lens is located at the center of the grid, chosen to be also the
origin of the coordinate system. The grid has a length of \(4 \time 4
\, \tE\) to cover the relevant area in which multiple images occur,
i.e., it covers the region within $2\theta_{\rm E}$ from the lens center. 

The SOR involves the calculation of a weighted average between the
previous iterate \(\ptilde_{i,k}^{(m-1)}\) and the computed
Gau\ss-Seidel iterate \(\tilde{\Psi}_{i,k}^{(m)}\) successively for
each component 
\begin{equation}
	\ptilde_{i,k}^{(m)} = \omega \, \tilde{\Psi}_{i,k}^{(m)} \, + \, ( 1 - \omega ) \, \ptilde_{i,k}^{(m-1)}\;,
\end{equation}
where \(\ptilde_{i,k}^{(m)}\) is the value of \ptilde \ for the grid
point \((i,k)\) in iteration \(m\), and \(\omega\) is the
extrapolation parameter. The parameter \(\omega\) is chosen such that
it accelerates the rate of convergence of the iterative variable to
the solution; in this work 
\begin{equation}
	\omega = \frac{2}{1 + \pi / (\N - 1) }\;,
\end{equation}
is applied, where \(\N \times \N\) \ is the total number of grid
points. Initially, all \(\ptilde_{i,k}\) are set to zero. In each
iteration \(m\), the Gau\ss-Seidel iterate
\(\tilde{\Psi}_{i,k}^{(m)}\) is calculated as follows (a
fourth-order accurate finite differencing is used) 
\begin{align}
	\tilde{\Psi}_{i,k}^{(m+1)} = &-\frac{1}{60} \, \Bigl( \ptilde_{i+2,k}^{(m)} + \ptilde_{i-2,k}^{(m)} + \ptilde_{i,k+2}^{(m)} + \ptilde_{i,k-2}^{(m)} \Bigl) \nonumber\\
	&+ \frac{16}{60} \, \Bigl( \ptilde_{i+1,k}^{(m)} + \ptilde_{i-1,k}^{(m)} + \ptilde_{i,k+1}^{(m)} + \ptilde_{i,k-1}^{(m)} \Bigl) \nonumber\\
	&- \frac{12}{60} h^2 \, \bigl[ \nabla \cdot \ahat \bigl]_{i,k}\;,
	\elabel{gaussseidel}
\end{align}
where \( h \) is the spacing of grid points. The divergence of  \ahat \ is calculated with fourth-order accurate finite differencing method for each grid point, and for points on the boundary of
the grid and the neighboring row and column, a second-order accurate
finite differencing scheme is employed.
Convergence is reached when two requirements are met: (i) at least \(
40 \, \N \) iterations have been made, and (ii) the maximum difference
\((\ptilde^{(m)}_{i,k} - \ptilde^{(m-1)}_{i,k})_\mathrm{max}\) between
two iterations increases. Typically, slightly more than \( 40 \, \N \)
are needed to reach convergence. If the process converges, the values
of \ptilde \ at the four corners are calculated by extrapolation. 

We consider that the typical accuracy on the image position of
observed lens systems is of the order \( 5 \, \mathrm{mas}\),
implying that $\eacc$ in Eq.\ts(\ref{eq:prec}) should be of the same
order (this choice will be discussed in
Sect.\ts\ref{sc:Sc4}). Thus, the numerical error of our method has to be
well below \(1 \, \mathrm{mas} \approx 10^{-3} \, \tE\) for typical
galaxy scale lenses which is quite stringent. Increasing the grid size
yields a strong increase in computational
time, which scales roughly as ${\cal N}^3$. 
Therefore, we added an extrapolation method to the standard SOR to
increase accuracy with a more reasonable increase in computational
time. The principle of our extrapolation scheme is displayed
\begin{figure}
	\begin{center}
		\includegraphics[width=.3\textwidth]{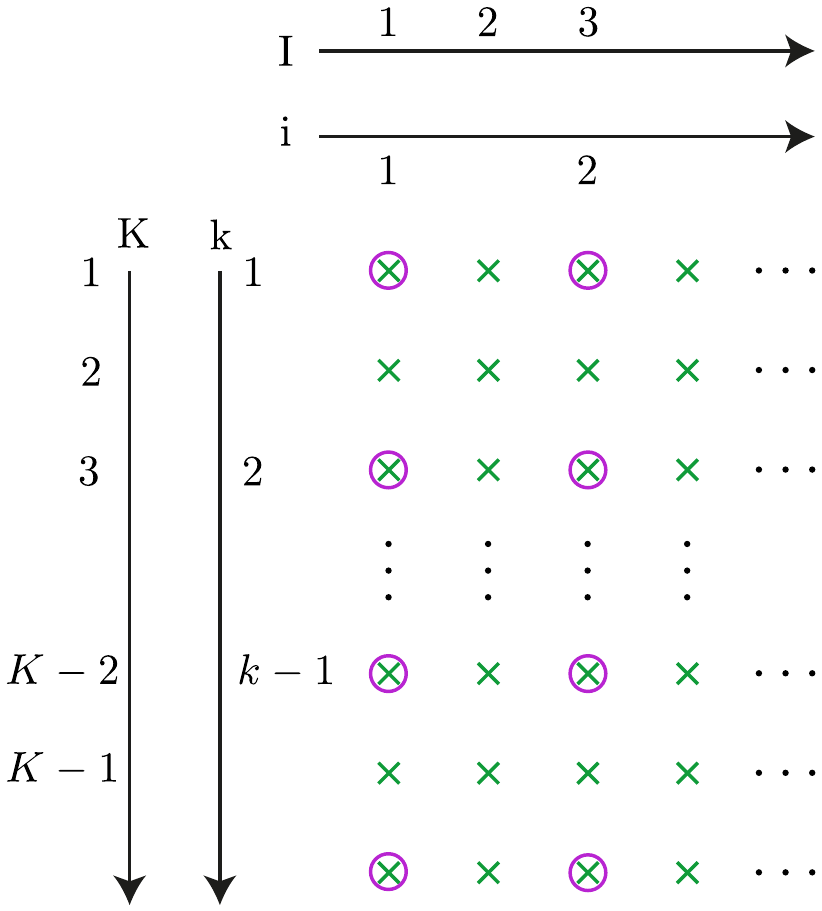}
		\\[10pt]
		\includegraphics[width=.3\textwidth]{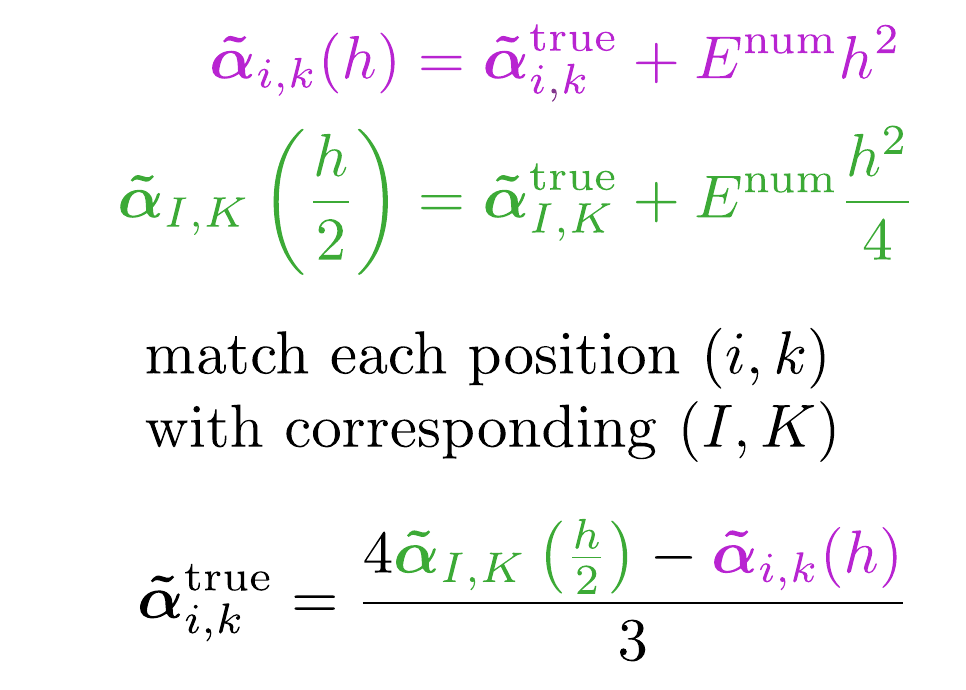}
	\end{center}
	\caption{An illustration of the extrapolation method used in the SOR method (Sect. \ref{sec:solv_neumann}) to calculate \atilde \ is shown: Based on the calculation of \atilde \ on two grids with indices \((i,k)\) and \((I,K)\), we can retrieve \(\atilde^\mathrm{true}\) with a minimum accuracy \(\Delta \alpha \) using the scheme described in the figure.}
	\label{pic:xtrapol}
\end{figure}
in Fig.~\ref{pic:xtrapol} and is based on the observation that the
error \( | \ealpha | \) of the computed value \( \atilde ( h ) \)
scales as $h^2\propto {\cal N}^{-2}$. This can be seen in the top
panel of Fig.~\ref{pic:error} where we applied our numerical scheme to
the case of a non-singular isothermal sphere, i.e., where the true
solution is known analytically. In this case, the deflection law \ahat
\ is a pure gradient field, and thus \( \avec = \ahat = \atilde \).
Using this scaling behavior we can extrapolate to the true deflection
\( \atilde^\mathrm{true} \), which would be obtained in the limit
$h\to 0$, for every grid point
\begin{equation}
	\atilde_{i,k} ( h ) = \atilde_{i,k}^\mathrm{true} + E^\mathrm{num}_{i,k} (h)^2\;,
\end{equation}
where \(E^\mathrm{num}\) is the numerical error.\footnote{Note that
  this extrapolation has to be carried out with the deflection angle,
  not with the potential, since the latter is determined only up to an
  additive constant -- which may depend on the iteration step $m$ and
  the number of grid points.} However, the
asymptotic deflection \( \atilde^\mathrm{true} \) and the value of
the numerical error \(E^\mathrm{num}\) are unknown in general. We can determine
the two unknowns by calculating \atilde \ for two different values of
\( h \), i.e., for different \N. Hence, we calculate \atilde \ on two
grids, of \( \N_1 = 2 N \) and \( \N_2 = N \) points. The coordinates
of the first and second grid are denoted respectively with indices \(
(I,K) \) and \( (i,k) \) and we have to match every grid point \(
(i,k) \) with its corresponding position \( (I,K) \). Then we can
obtain the true value \( \atilde^\mathrm{true} \)
\begin{equation}
	\atilde^\mathrm{true}_{i,k} = \frac{ 4 \atilde_{I,K} \left( \frac{h}{2} \right) - \atilde_{i,k} (h) }{3}\;,
\end{equation}
as indicated in Fig.~\ref{pic:xtrapol}.

Incorporating this extrapolation method in the code decreases the
numerical error for the grid point numbers that are used (\(\N \sim
400\)) roughly by a factor of \(10^3\), as can be seen in the lower
panel of Fig.~\ref{pic:error}, which also shows that the numerical
error with this extrapolation scheme decreases much faster with ${\cal
  N}$ than without.
\begin{figure}[htbp]
	\begin{center}
		\includegraphics[width=.45\textwidth]{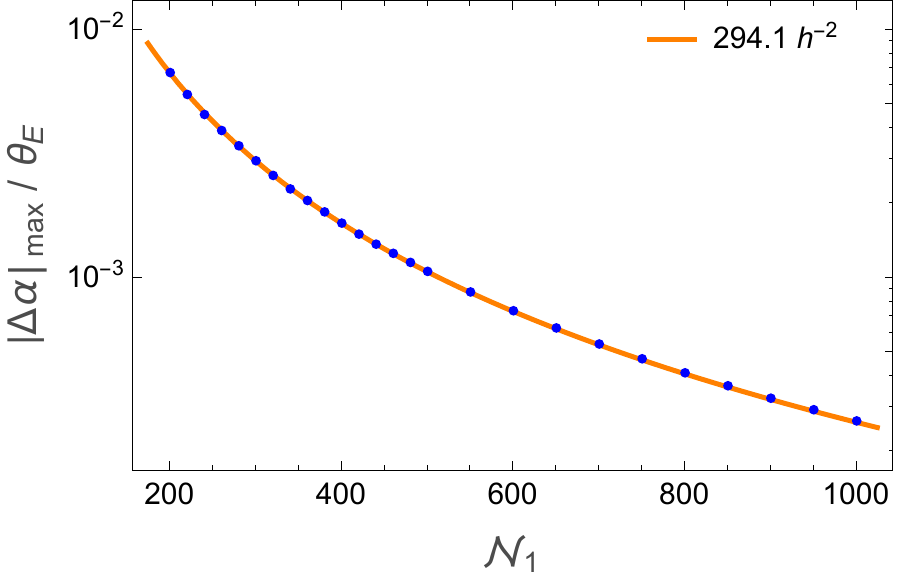}
		\includegraphics[width=.45\textwidth]{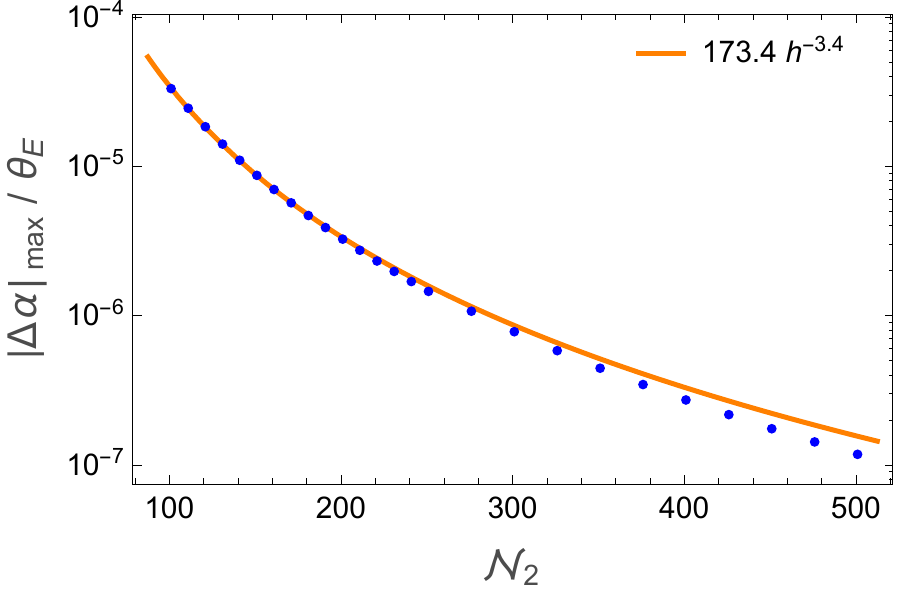}
	\end{center}
	\caption{Maximum difference of \( | \ahat - \atilde | = |
          \ealpha | \) for a non-singular isothermal sphere with core
          radius \(\tc = 0.1 \, \tE \) as a function of the number of
          grid points \mbox{number \N} used in the numerical
          solution. Blue dots are the numerical results, whereas the
          curves present power-law fits to these points with \( h \) being the spacing of grid points. In the top
          panel, the results are shown for the `standard' method,
          where the numerical error scales as ${\cal
            N}^{-2}$. Incorporating the extrapolation scheme, the
          numerical error decreases much faster with the number of
          grid point, as can be seen in the lower panel (note the different scale for the \(y\)-axis in the upper and lower panel). For the
          typical values used in the papers (${\cal N}\sim 400$), a
          gain in accuracy by three orders of magnitude is obtained
          with extrapolation with only a modest increase of
          computational cost ($\sim 25\%$). Since extrapolation
          includes calculating 
          \atilde \ twice with grid points \(2N\) and \(N\), values
          are only shown up to \(\N_2 = 500\) which correspond to \( \N_1 =
          1000 \) in the graph above.}
	\label{pic:error}
\end{figure}
The largest numerical deviation \((\Delta \alpha)_\mathrm{max}\) can
be found near the center of the grid. This is as expected, since the
deviation from \( \atilde^\mathrm{true} \) depends on higher-order
derivatives, which for the chosen lens model are largest near the
center. However, multiple images near the center of the lens are
usually highly demagnified and rarely observable for galaxies as
lenses \citep[see, e.g.,][]{Hezaveh15,Winn04} and are therefore not
relevant. 

We have also tested the accuracy of the numerical implementation. One
method is to check whether \( \nabla \cdot \ahat (\tvec) = \nabla
\cdot \atilde (\tvec) \) is valid for the whole grid for any
deformation function \(\bhat (\bvec)\). In all our calculations,
deviations \( | \khat - \ktilde | \) are always smaller than
\(10^{-4}\) if the corner regions, i.e.,
\( \tvec \ge 2 \, \tE\), are excluded from our analysis. Thus, we only
consider the behavior of the mass profile in the circular region \(
|\tvec|/\tE < 2 \), where numerical errors in $| \ahat - \atilde |$
do not exceed \(10^{-6} \, \tE\).
\subsection{Solution by means of a Green's function}
\label{sec:solv_green}
A different approach is to find a solution of Eq.\ts (\ref{eq:vNP}) by
means of a Green's function. For that, we make use of Green's second
theorem, considering a function $h(\vc\theta)$,
\begin{equation}
	\int_{\cal U} \d^2\theta\;\eck{\tilde\psi\,\nabla^2 h-h\,\nabla^2 \tilde\psi} =\int_{\partial \cal U}\d s\; \eck{\tilde\psi\,
	\nabla h\cdot \vc n- h\,\nabla\tilde\psi\cdot \vc n}\;.
	\elabel{GreenT}
\end{equation}
We choose the function $h(\vc\theta)\equiv H(\vc\vt;\vc\theta)$ depending on the vector $\vc\vt$ such that it obeys the following equations:
\begin{eqnarray}
	\nabla_\theta^2 H(\vc\vt;\vc\theta)=\delta(\vc\theta-\vc\vt)-\frac{1}{A}\;\;& {\rm for} &\;\; \vc\theta\in {\cal U}\;, \nonumber\\
	\nabla_\theta H(\vc\vt;\vc\theta)\cdot \vc n =0 \;\; &{\rm for} &\;\; \vc\theta\in\partial {\cal U}\;, 
	\elabel{Greens}
\end{eqnarray}
where $A$ is the area of $\cal U$, and $\vc\vt$ is a point within
$\cal U$. The term $1/A$ in Eq.\ts (\ref{eq:Greens}) is needed to
satisfy Green's divergence theorem applied to the vector field $\nabla
h$, which requires  
\begin{equation}
	\int_{\cal U} \d^2\theta\;\nabla^2 h =
	\int_{\partial \cal U}\d s\;\nabla h\cdot \vc n\;;
\end{equation}
the conditions (\ref{eq:Greens}) set both side of this equation to
zero. With (\ref{eq:Greens}), Eq.\ts (\ref{eq:GreenT}) simplifies to
\begin{eqnarray}
	\tilde\psi(\vc\vt)\!\!\!&=&\!\!\!\ave{\tilde\psi}+\int_{\cal U}\!\! \d^2\theta\; H(\vc\vt;\vc\theta)\,\nabla^2 \tilde\psi -\int_{\partial \cal U}\!\!\!\d s\;H(\vc\vt;\vc\theta)\,\nabla\tilde\psi\cdot \vc n  \nonumber\\
	\!\!\!&=&\!\!\!\ave{\tilde\psi}+2\int_{\cal U} \!\! \d^2\theta\; H(\vc\vt;\vc\theta)\,\hat\kappa(\vc\theta) -\int_{\partial \cal U}\!\!\!\d s\;H(\vc\vt;\vc\theta)\,\hat{\vc\alpha}\cdot \vc n\;,
	\elabel{vNSol}
\end{eqnarray}
where $\ave{\tilde\psi}$ is the average of $\tilde\psi$ on ${\cal U}$,
and we used Eq.\ts (\ref{eq:vNP}) in the last step. We note that the integral 
\begin{equation}
	f(\vc\theta)=\int_{\cal U}\d^2\vt\;H(\vc\vt;\vc\theta)
\end{equation}
is a constant, since $\nabla^2 f(\vc\theta) =0$ and $\vc n\cdot \nabla
f=0$ on the boundary of $\cal U$. Therefore, if we integrate Eq.\ts
(\ref{eq:vNSol}) over $\cal U$, the two integrals on the
r.h.s. compensate each other, due to the divergence theorem, so that
this solution is consistent.

Whereas for a general region $\cal U$ it will be difficult to obtain a
solution of Eq.\ts (\ref{eq:Greens}) for $H(\vc\vt;\vc\theta)$, such a
solution is analytically known if $\cal U$ is a circle of radius
$R$. In this case,
\begin{eqnarray}
	H(\vc\vt;\vc\theta)&=&\frac{1}{4\pi} \eck{\ln\frac{\abs{\vc\vt-\vc\theta}^2}{R^2} +\ln\rund{1-\frac{2\vc\vt\cdot \vc\theta}{R^2}+\frac{|\vc\vt|^2|\vc\theta|^2}{R^4}}}\nonumber \\
	&-&\frac{|\vc\vt|^2+|\vc\theta|^2}{4\pi R^2}\;,
	\elabel{Hcirc}
\end{eqnarray}
which has the properties that
\begin{eqnarray}
	&&\nabla^2_\vt H(\vc\vt;\vc\theta) =\delta(\vc\vt-\vc\theta)-\frac{1}{\pi R^2} = \nabla^2_\theta H(\vc\vt;\vc\theta)\;\;\hbox{for}\;\;\vc\vt, \vc\theta\in \cal U \;,\nonumber \\
	&&\nabla_\theta H(\vc\vt;\vc\theta) \cdot \vc n(\vc\theta)=0 \;\; \hbox{for}\;\;
	\vc\theta\in \partial\cal U\;. \nonumber
\end{eqnarray}
Hence, Eq.\ts (\ref{eq:Hcirc}) indeed satisfies the conditions
(\ref{eq:Greens}). 

With this explicit solution, we can now calculate the deflection angle
corresponding to the potential $\tilde\psi$,
$\tilde{\vc\alpha}=\nabla\tilde\psi$, by obtaining the gradient of
$H$,
\begin{equation}
	\nabla_\vt H(\vc\vt;\vc\theta)=\frac{1}{2\pi} \rund{\frac{\vc\vt-\vc\theta}{|\vc\vt-\vc\theta |^2}-\frac{\vc\vt}{R^2} +\frac{  |\vc\theta|^2\vc\vt- R^2\vc\theta}{\rund{R^4-2 R^2 \vc\vt\cdot\vc\theta+|\vc\vt|^2|\vc\theta|^2}}} \;.
	\elabel{nabH}
\end{equation}
Then,
\begin{equation}
	\atilde (\vc\vt)  = 2\int_{\cal U} \!\! \d^2\theta\; \nabla_\vt H(\vc\vt;\vc\theta)\,\hat\kappa(\vc\theta) -\int_{\partial \cal U}\!\!\!\d s\; \nabla_\vt H(\vc\vt;\vc\theta)\,\hat{\vc\alpha}\cdot \vc n\;,
	\elabel{Hatilde}
\end{equation}
where we have to deal with a pole in the first term of
Eq.\ts (\ref{eq:nabH}). Using a conformal mapping as described in Appendix
\ref{sc:ApA}, we can handle this pole numerically. In the third term
the pole lies outside the circle and since \( \vc\vt \in \U
\) there is no pole. However, if \tvec \ is on the circle (as occurs
in the line integral in (\ref{eq:Hatilde})), the third term can
become quite large; hence, for points $\vc\vt$ near the boundary,
special care is needed to obtain an accurate solution.

This Green's function approach has several advantages over using a SOR
for a grid. First, the region on which the von Neumann problem is
defined can be chosen as a circle, instead of a rectangle for the SOR
method. Second, the solution by means of the Green's function yields
higher accuracy. The reason for this is that the
limited accuracy in finite differencing does not occur here. Third, if
one is interested in the deflection only at a few points, this can be
calculated much faster than with the SOR which necessarily calculated
the solution over the whole region.

\subsection{Interpretation}
The expression (\ref{eq:vNSol}) for the deflection potential, or the
expression (\ref{eq:Hatilde}) for the deflection angle, contains quite
a number of terms. In order to obtain a better understanding of the
various terms, we consider again the case where the deflection angle
$\hat{\vc\alpha}$ is a pure gradient field, in which case
$\tilde{\vc\alpha}=\hat{\vc\alpha}\equiv \vc\alpha$. Then the
deflection angle at a point $\vc\vartheta\in {\cal U}$ can be
decomposed into a deflection $\vc\alpha_{\rm in}$ which is caused by
matter inside ${\cal U}$, and one due to matter outside ${\cal U}$,
denoted by $\vc\alpha_{\rm out}$. Thus we expect that
\begin{equation}
	\vc\alpha(\vc\vt)=\vc\alpha_{\rm in}(\vc\vt)+\vc\alpha_{\rm out}(\vc\vt) = \frac{1}{\pi}\int_{\cal U}\d^2\theta\;\kappa(\vc\theta)\, \frac{\vc\vt-\vc\theta}{|\vc\vt-\vc\theta |^2} +\vc\alpha_{\rm out}(\vc\vt) \;.
	\elabel{A1}
\end{equation}
Comparing the last equation (\ref{eq:A1}) to (\ref{eq:Hatilde}), we find that 
\begin{equation}
	\vc\alpha(\vc\vt)=\vc\alpha_{\rm in}(\vc\vt)+\vc A(\vc\vt) -\vc B_{\rm in}(\vc\vt)-\vc B_{\rm out}(\vc\vt)\;,
	\elabel{A3}
\end{equation}
where 
\begin{eqnarray}
	\vc A(\vc\vt)&=&\int_{\cal U} \!\! \d^2\theta\;\kappa(\vc\theta)\, \rund{2\nabla_\vt H(\vc\vt;\vc\theta)-\frac{1}{\pi} \frac{\vc\vt-\vc\theta}{|\vc\vt-\vc\theta |^2}}\;, \nonumber \\
	\vc B_{\rm in, out}(\vc\vt)&=&\int_{\partial \cal U}\!\!\!\d s\;\nabla_\vt H(\vc\vt;\vc\theta)\;{\vc\alpha}_{\rm in, out}\cdot \vc n\;,
	\elabel{A4.1}
\end{eqnarray}
where we split the deflection angle on the boundary into terms due to
matter inside and outside ${\cal U}$. Both of the terms $\vc A$ and
$\vc B_{\rm in}$ are due to matter inside ${\cal U}$, whose deflection
is covered entirely by the first term $\vc\alpha_{\rm in}$, so that we
expect that
\begin{equation}
	\vc A(\vc\vt)=\vc B_{\rm in}(\vc\vt)\;.
	\elabel{A5}
\end{equation}
In Appendix B we show explicitly that this relation holds for the case
of a circular region for which $H$ is given by Eq.\ts
(\ref{eq:nabH}). Hence, Eq.\ts (\ref{eq:A3}) then provides a clean
separation of the deflection angle coming from the inner mass
distribution ($\vc\alpha_{\rm in}$) and that coming from matter
outside $\cal U$, given by $\vc B_{\rm out}$. This relation may be of
practical relevance for the numerical calculation of the lensing
properties from a complicated mass distribution, for which the lensing
quantities are only needed inside a limited region. Instead of
calculating, for every point inside $\cal U$, a two-dimensional
integral of the surface mass density $\kappa$ over the whole lens
plane, one can proceed as follows: First, one can reduce the
integration range over the region $\cal U$ to get the contribution
$\vc\alpha_{\rm in}$. Second, one can calculate the contribution
$\vc\alpha_{\rm out}$ for points on the boundary by integrating over
the outer region of the lens in terms of a two-dimensional
integral. Third, the contribution $\vc\alpha_{\rm out}$ for points
inside the region ${\cal U}$ can then be obtained by a one-dimensional
integration over the boundary curve.

In general, if $\vc\alpha$ is given on the boundary, it contains
contributions from both the inner and the outer part. In other words,
the split of $\vc B$ into $\vc B_{\rm in}$ and $\vc B_{\rm out}$ is
not provided in that case. The term $\vc A$ then compensates for the
contribution $\vc B_{\rm in}$ of $\vc B$.
\section{\llabel{Sc4}Illustrative example - a quadrupole lens and an
  isotropic SPT }
Our goal is to find criteria allowing us to assess whether an
SPT-transformed deflection law is valid (i.e. deviates from a gradient
field by less than \eacc), using the methods explained in the previous
section. Thus, we set an upper limit on how much the transformed
deflection law \ahat \ is allowed to differ from its closest curl-free
approximation \atilde \ before leading to a non-negligible shift of
the lensed images. Since observed lens systems are usually fit by
simple mass models with only a small number of free parameters, we do
not expect the fit to be perfect. We always have to deal with
observational uncertainties as well as the presence of substructure
\citep{Dandan10,Brad04,KochDal04,MS98} and line-of-sight
inhomogeneities \citep{Dandan12,Metcalf05}. Therefore, we cannot
reproduce observed positions better than a few milliarcseconds with a
smooth mass model. Hence, as long as \(| \ealpha (\tvec) |\) is less
than the smallest angular scale on which modeling with a smooth mass
model is still meaningful, differences are of no practical relevance
(SS14).

We need to choose a lens model to explore how seriously the SPT may
affect lens modeling. First, we consider a situation similar to SS14,
namely a quadrupole lens
with external shear \gp
\begin{equation}
	\avec = \kbar (|\tvec|) \, \tvec - \begin{pmatrix} \gp & 0 \\ 0 & -\gp \end{pmatrix} \ \tvec
	\elabel{quadlens}
\end{equation}
which is deformed by an SPT corresponding to a radial stretching, as in
Eq.\ts (\ref{eq:radstretch}). Specifically, we choose
\begin{equation}
	\bhat (\bvec) = \left( 1 + \frac{f_2}{2 \tE^2} \beta^2 \right) \bvec \;.
	\elabel{deffct}
\end{equation}
This deformation function is the lowest-order expansion of more
general stretching functions, and its leading-order term is chosen
such as to not yield an MST, to cleanly separate the effect of the
MST from that of the more general SPT in this study.
Furthermore, we choose as specific lens model 
a non-singular isothermal sphere (NIS), described by the mean
convergence profile
\begin{equation}
	\kbar = \tE \, \frac{1}{\sqrt{\tc^2 + \theta^2}}\;,
\end{equation}
where $\tc$ is the core radius. For the rest of this paper, we fix the
core to be $\tc=0.1\tE$. 

To get a quantitative estimate on how large deviations of
$\tilde{\vc\alpha}$ from $\hat{\vc\alpha}$ are tolerable before the
lensing properties of the SPT deviate markedly from the original lens
model, we take the Hubble Space Telescope (HST) as example. We
estimate that the highest astrometric accuracy that can be achieved
corresponds to about a tenth of a pixel in the ACS camera, or \(\Delta
\theta \approx 5 \, \mathrm{mas} \approx 5 \times 10^{-3} \, \tE\),
where the last expression accounts for the fact that the typical
Einstein radii of galaxy-scale lenses are of order one
arcsecond. Thus, if the solution $\tilde{\vc\alpha}$ satisfies the
condition (\ref{eq:prec}) with $\eps_{\rm acc}= 5\times 10^{-3}\,\tE$
over the region $|\vc\theta|\le 2\,\theta_{\rm E}$,
we call the corresponding SPT `allowed' or `valid'.

\subsection{Impact on the deflection law}
\begin{figure}[htbp]
	\centering
	\includegraphics[width=.49\textwidth]{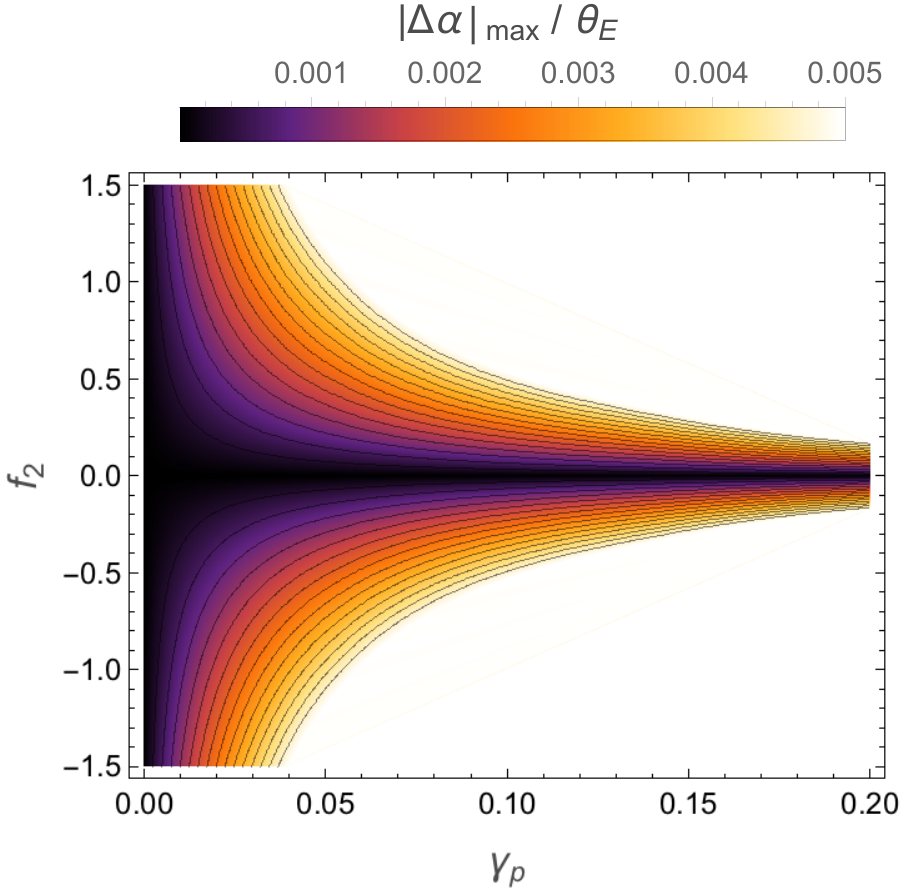}
	\caption{Values of \(|\ealpha|_\mathrm{max}\) are plotted
          against the parameters \(f_2\) from
          (\ref{eq:deffct}) and external shear strength \gp. The
          colored region indicates allowed pairs of parameters that
          fulfill the \(|\ealpha| < 5 \times 10^{-3} \,
          \tE\)-criterion. For obtaining this figure, we used the SOR method.}
	\label{pic:f2gp}
\end{figure}

The model we consider has two free parameters, the distortion
parameter $f_2$ in the SPT (\ref{eq:deffct}), and the strength $\gp$ of
the external shear.  We start with exploring this parameter space 
to find the combination that yield allowed transformations, using the
methods described in the previous section. In
Fig.~\ref{pic:f2gp}, we display the maximum deviation
$|\Delta\vc\alpha|_{\rm max}$ as a function of these two parameters.
It shows
a wide range of allowed parameter combinations, where the allowed
range of $f_2$ decreases with increasing external shear. The white
regions in Fig.~\ref{pic:f2gp} denotes parameter combinations where 
$|\Delta\vc\alpha|_{\rm max}> 0.005\,\tE$, and which are therefore not
allowed according to our accuracy criterion.

In SS14, we speculated that the curl of $\ahat$ may yield a good
indication for the deviation of the SPT-transformed deflection field
from a gradient field. In this case, the curl \( \khat_I = \nabla
\times \ahat \), which describes the asymmetric part of the Jacobian,
could be used as a proxy for $|\Delta\vc\alpha|$. For a quadrupole lens
of the form (\ref{eq:quadlens}) and the deformation law
(\ref{eq:deffct}), the curl $\khat_{\rm I}$ is given in Eq.\ts (42) of
SS14,

\begin{align}
	\ki &\approx - \frac{\gp}{2} f_2 \left( \frac{\theta}{\tE} \right)^2 \label{eq:kiSS14}\\
	&\times \left[\gp^2 - (1 - \kbar)(2\gm + 1 - \kbar) + 2 \gm \gp \cos(2\varphi)\right] \sin 2\varphi \;, \nonumber
\end{align}
where \(\theta, \varphi\) describe polar coordinates in the lens plane
and \( \gm(|\tvec|) = \kappa (|\tvec|) - \kbar (|\tvec|) \) is
the shear caused by the NIS lens.

\begin{figure}[htbp]
	\centering
	\includegraphics[width=.49\textwidth]{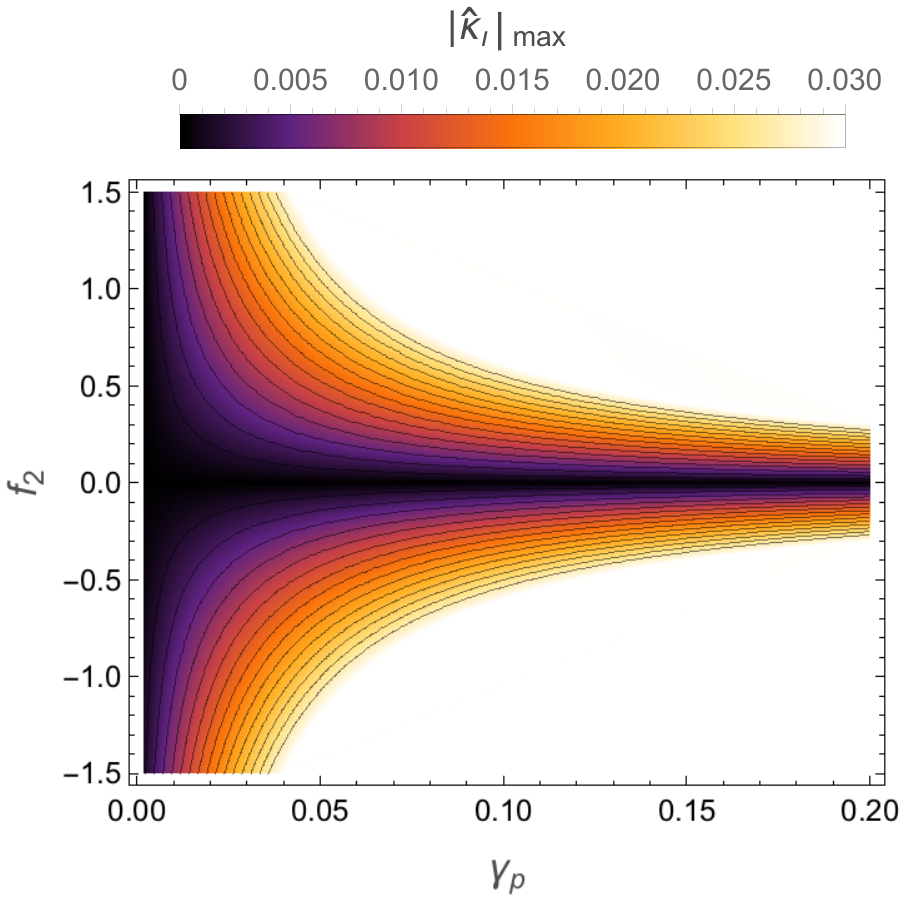}
	\caption{Values of \(|\ki|_\mathrm{max}\) are plotted against
          the parameters \(f_2\) from (\ref{eq:deffct}) and
          external shear strength \gp. The colored region indicates
          allowed pairs of parameter that were chosen such that they
          roughly correspond to \(|\ealpha| < 5 \times 10^{-3} \,
          \tE\).}
	\label{pic:f2gpkapI}
\end{figure}

\begin{figure}[htbp]
	\centering
	\includegraphics[width=.49\textwidth]{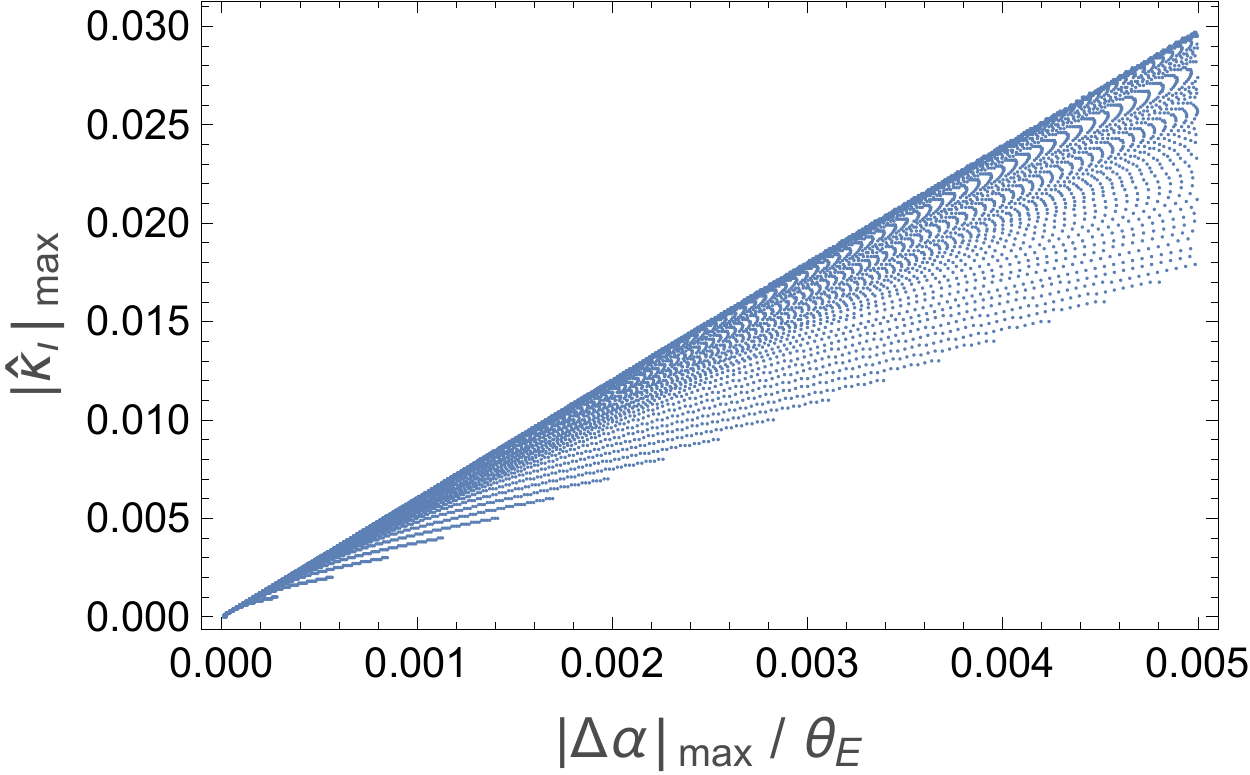}
	\caption{For every allowed combination \(f_2\) and \gp \ the
          values of \(|\ealpha|_\mathrm{max}\) \ (Fig.~\ref{pic:f2gp})
          are plotted against \(|\ki|_\mathrm{max}\)
          (Fig.~\ref{pic:f2gpkapI}). A clear correlation between these
        two quantities can be seen.}
	\label{pic:correl}
\end{figure}

Fig.~\ref{pic:f2gpkapI} shows the maximum of \ki \ as a function of external shear \gp \ and deformation \enquote{strength} \(f_2\),
which indeed is very similar to Fig.~\ref{pic:f2gp}. The actual difference between those two approaches is seen in Fig.~\ref{pic:correl}. An approximately linear correlation
with an expected but modest scatter can be seen. In fact, from that
figure we obtain for our specific model that
\begin{equation}
0.16\, |\ki|_\mathrm{max} \lesssim {|\Delta\vc\alpha|_{\rm max}\over
  \tE} \lesssim  0.3\, |\ki|_\mathrm{max}  \;.
\elabel{inequ}
\end{equation}
For other models, the relation between $|\Delta\vc\alpha|_{\rm max}$
and $|\ki|_\mathrm{max}$ will be different; nevertheless, we see that
the curl of $\ahat$ indeed provides a useful indication for the
validity of an SPT, since calculating  $\ki$ is much easier then
obtaining the numerical solution for $\atilde$.

\begin{figure}[htbp]
	\centering
	\includegraphics[width=.45\textwidth]{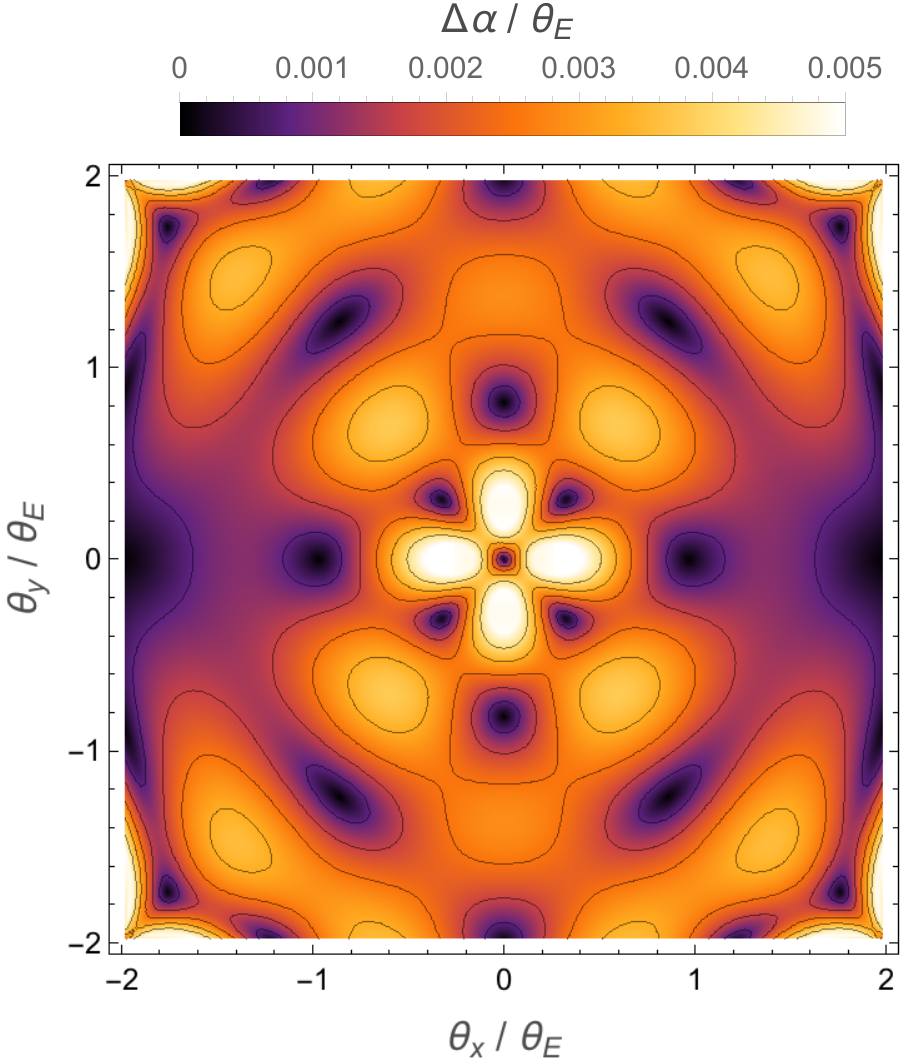}
	\caption{A map of \( | \ealpha (\tvec)| \) is shown for \(f_2 = 0.55\)
          and \(\gp = 0.1\). The strong changes in the corners, i.e.~\(\theta > 2 \, \tE\), are biased by large numerical uncertainty and should
          be neglected.}
	\label{pic:alphadiff}
\end{figure}

Figure \ref{pic:alphadiff} illustrates how a specific deflection law in
a region \( |\tvec| \le 2 \,\tE \) is affected by an SPT. It shows \(|
\ealpha (\tvec) |\) for a quadrupole lens with external shear \(\gp =
0.1\) and deformation strength \(f_2 = 0.55\), which is
the highest allowed for this value of the external shear strength and
thus is expected to show
the largest deviations \khat \ compared to the original mass
profile. The figure shows that the largest
deviations occur at an angle of \(45^\circ\) with respect to the
external shear. This pattern, which is shown for one specific pair of
\(f_2\) and \gp, is qualitatively the same for all
\(f_2\)-\gp-combinations.
\subsection{Implications for the convergence}
\begin{figure}[htbp]
	\centering
	\includegraphics[width=.45\textwidth]{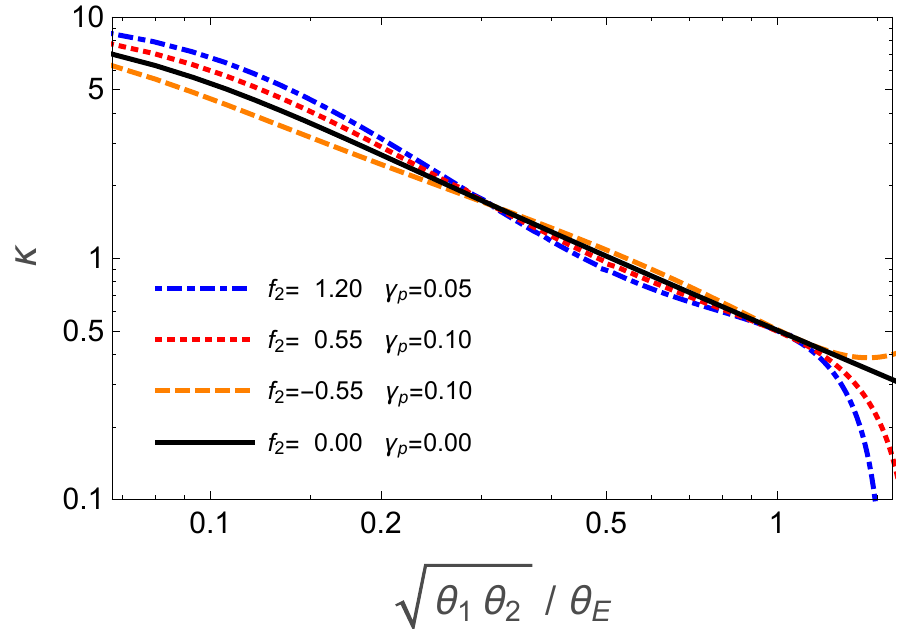}
	\\[10pt]
	\includegraphics[width=.45\textwidth]{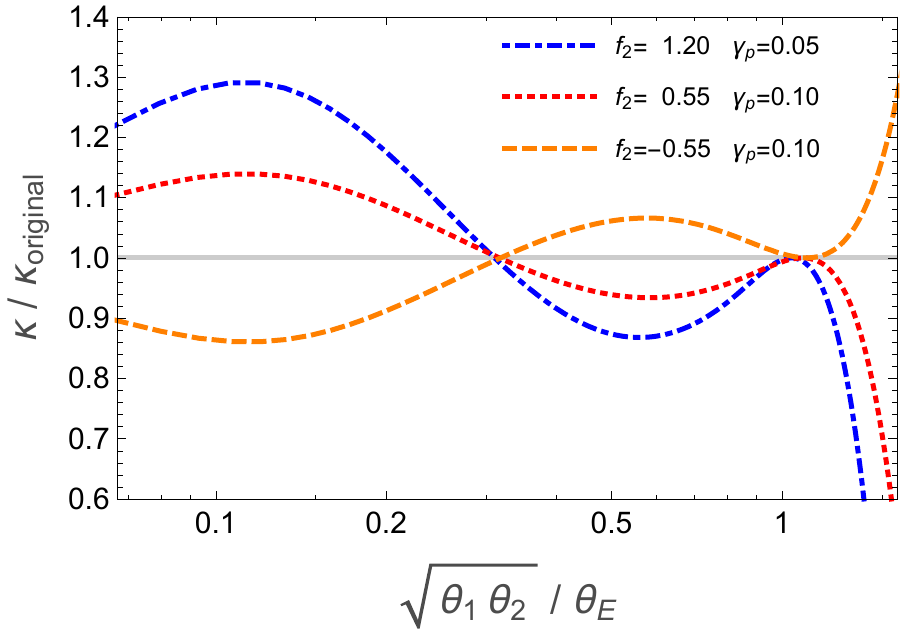}
	\caption{The upper panel shows the mass profile of the
          original NIS lens (solid curve), and that of three
          SPT-transformed lenses, with parameters $f_2$ and $\gp$
          indicated by the labels. For all of these three models,
          \( \ealpha_\mathrm{max} \approx
          \eacc = 5 \times 10^{-3} \, \tE \). Since the transformed
          mass distributions have a finite ellipticity, the density is
          plotted as a function of the geometric mean of the major and
          minor semi-axis of the best-fitting ellipse to an isodensity
          contour, except for the case with negative $f_2$, for which
          the outer isodensity contours are not closing around the
          lens center; in this special case, the x-axis corresponds to
          the \(\theta_1\)-axis. The convergence
          changes up to \(28\%\) for radii smaller than \(1\,\tE\),
          radii larger than that show a significantly smaller
          convergence for a positive \(f_2\). Negative \(f_2\) show an
          essentially mirrored behavior compared to positive
          \(f_2\). This leads to convergence \khat \ that may not
          decrease
          monotonically. The lower panel shows
          the ratio between transformed and original mass profile.}
	\label{pic:kapradial}
\end{figure}

We show in Fig.~\ref{pic:kapradial} the comparison between original
($\kappa$) and SPT-transformed mass distribution ($\hat\kappa$) for
three different allowed pairs of parameters, \(f_2= 0.55\) and \(\gp=0.1\)
(the same combination of parameters as in Fig.~\ref{pic:alphadiff}),
\(f_2=-0.55\) and \(\gp=0.1\), and \(f_2 = 1.2\) and \(\gp=0.05\). The
lower panel of Fig.~\ref{pic:kapradial} shows
the change of the radial profile as \(\khat/\kappa_\mathrm{original}\).

The divergence of \ahat \ (i.e.~$\nabla\cdot\ahat=2\hat\kappa$), was
calculated analytically in SS14 (see their Eq.\ts 41) and can be used
to compare our numerical results to the analytic solution. Specialized
to our case, it reads
\begin{align}
	\khat = &\ \kappa_\mathrm{NIS} + \frac{f_2}{2} \left( \frac{\theta}{\tE} \right)^2 \times \nonumber\\\label{eq:khatSS14}
	&\ \biggl( \ \gamma_\mathrm{m} \Bigl[ 2 \gamma_\mathrm{p}^2 + 3 \bigl( 1 - \kbar \bigl)^2 \Bigl] \ - \ 2 \bigl( 1 - \kbar \bigl) \Bigl[ \bigl( 1 - \kbar \bigl)^2 + 2 \gamma_\mathrm{p}^2 \Bigl] \biggl. \\
	&\ + \Bigl[ 5 \gamma_\mathrm{p} \bigl( 1 - \kbar \bigl)^2 - 6 \gamma_\mathrm{p} \gamma_\mathrm{m} \bigl( 1 - \kbar \bigl) + \gamma_\mathrm{p}^3 \Bigl] \, \cos 2 \varphi + \gamma_\mathrm{p}^2 \gamma_\mathrm{m} \, \cos 4 \varphi \biggl. \biggl) \ . \nonumber
\end{align}
where again \(\theta, \varphi\) describe polar coordinates in the lens
plane. The change $\Delta\kappa=\khat- \kappa_\mathrm{NIS}$ is
proportional to the stretching parameter $f_2$, so that
$\Delta\kappa(-f_2)=-\Delta\kappa(f_2)$. This behavior can be seen in
Fig.\ts\ref{pic:kapradial}. Indeed, we checked that all numerically
obtained deflection angles $\atilde$ are such that their corresponding
surface mass densities agree with the analytical prediction
(\ref{eq:khatSS14}). For example, the numerical result for the
parameter combination \(\gp = 0.1\) and \(f_2 = 0.55\) deviates by
less than \(3 \times 10^{-3}\) from the analytical solution.

As seen from Eq.\ts (\ref{eq:khatSS14}), the resulting mass
distribution $\khat$ is no longer axi-symmetric, but that 
isodensity contours are nearly elliptical (i.e., a factor
proportional to \(\cos(2\phi)\)) with a small boxiness (i.e., a factor
proportional to \(\cos(4\phi)\)). Hence, we define the distance from
the center generally as the geometric mean \(\sqrt{\theta_1
  \theta_2}\) using the 1- and 2-axis of the elliptical isodensity
contours. However, for sufficiently negative $f_2$, the outer
isodensity contours are no longer concentric, i.e., they are not
closed curves around the center of the lens. In addition, for large
negative values of $f_2$, the radial profile can become non-monotonic.
We consider such a
behavior as non-physical, i.e., such resulting models will be
irrelevant in practice.

\begin{figure}[htbp]
	\centering
	\includegraphics[width=.45\textwidth]{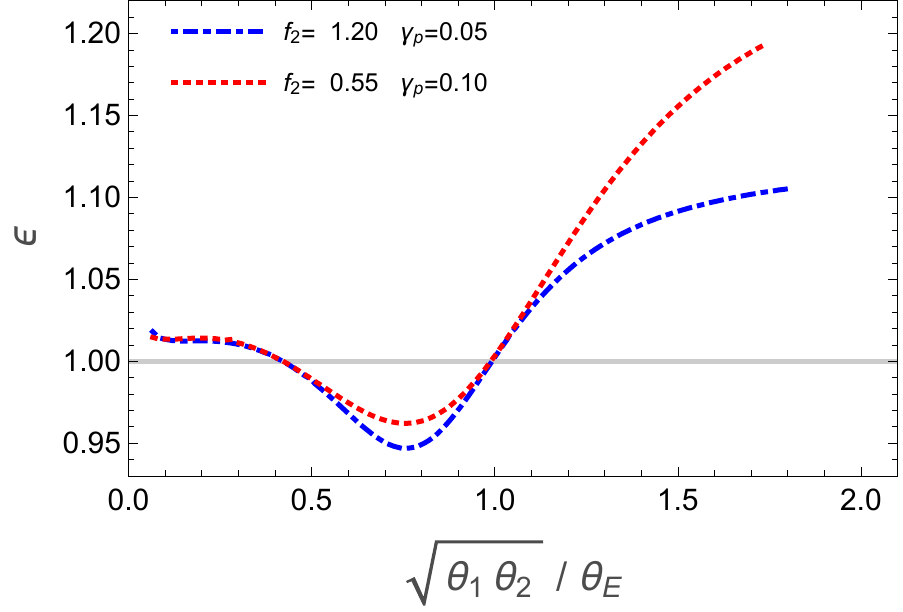}
	\caption{The radial dependence of the axis ratio
          $\epsilon$. In the unperturbed case the isodensity contours
          are circular, i.e., \(\epsilon (\sqrt{\theta_1\theta_2}) =
          1\). The SPTed mass distribution shows for radii
          \(\sqrt{\theta_1\theta_2} < 1 \, \tE\) deviations of up
          to \(5\%\) from circularity, whereas for larger radii the
          deviations can be up to \(20\%\). The convergence map for
          \(f_2=-0.55\) and \(\gp = 0.1\) does not show concentric
          isodensity contours for radii larger than \(1.3 \, \tE\) and
          therefore no ellipticity as a function radius can be
          determined.}
	\label{pic:kapell}
\end{figure}

The ellipticity of the transformed mass profiles is non-negligible as
shown in Fig.~\ref{pic:kapell} where \(\epsilon\), defined as the axis
ratio 1- over 2-axis, is plotted as a function of radius.

Integrating the analytic representation (\ref{eq:khatSS14}) of \khat \
up to \(1\,\tE\) it can be shown that the mass enclosed within the Einstein
radius {\it of the original lens}
is conserved, independent of the chosen mass profile \(\kappa
(\theta)\) (see Appendix \ref{sec:masscons}).
%
%
\section{\llabel{Sc5}Characterization of the modified mass distribution}
The SPT leads to a modified deflection angle of the lens which yields
exactly the same astrometric and photometric observational
properties as the original mass distribution. For those modified
profiles $\hat{\vc\alpha}$ for which a deflection potential
$\tilde\psi$ can be found such that the differences between the
corresponding \ealpha \ is sufficiently small, the modified surface
mass density $\hat\kappa$ provides a viable alternative to the
original mass model $\kappa$ of the lens. In this section we want to
consider a diagnostic for the change of the mass profile, both
regarding the radial slope and the angular structure of the
lens. Since the strong lensing properties of the lens can only be
probed in the inner part of the mass distribution, we will apply these
diagnostics only to those regions where multiple images can occur,
i.e., $|\vc\theta|\lesssim 2 \theta_{\rm E}$. 
\subsection{Radial mass profile}
The SPT changes the radial mass profile of the lens. We consider
situations in which the original lens is described by a
\enquote{simple} mass distribution, i.e., an NIS. Combined with a
\enquote{mild} SPT the resulting $\hat\kappa$ remains simple, e.g.,
still shows closed, concentric isodensity contours. 

The mass-sheet transformation is a special case of the SPT, and it is
well known that the MST changes the radial profile of the lens. In
order to highlight the new feature of the SPT not contained in the
MST, we aim at a measure for the radial profile which is invariant
under the MST. The MST transforms all derivatives of $\kappa$ by a
constant factor $\lambda$, hence it leaves the ratio of
derivatives unchanged. Consequently, one possible
  diagnostic for the effect of the SPT is the radial profile of such
ratios, e.g., $\ave{\kappa}''/\ave{\kappa}'$.

In particular, if the original mass profile is a power law,
$\ave{\kappa}(\theta) \propto \theta^{-s}$, then we have $\theta\,
\ave{\kappa}''/\ave{\kappa}' = -(s+1)$; hence, any deviation from this
constant value indicates the effect of the SPT on the modified mass
profile $\hat\kappa$. However, if there is no analytical expression of
$\kappa$ and $\hat{\kappa}$, the ratio of derivatives is very
sensitive to numerical noise, and therefore of little practical
interest. We therefore consider hereafter alternative tests.

\begin{figure}[htbp]
	\centering
	\includegraphics[width=.5\textwidth]{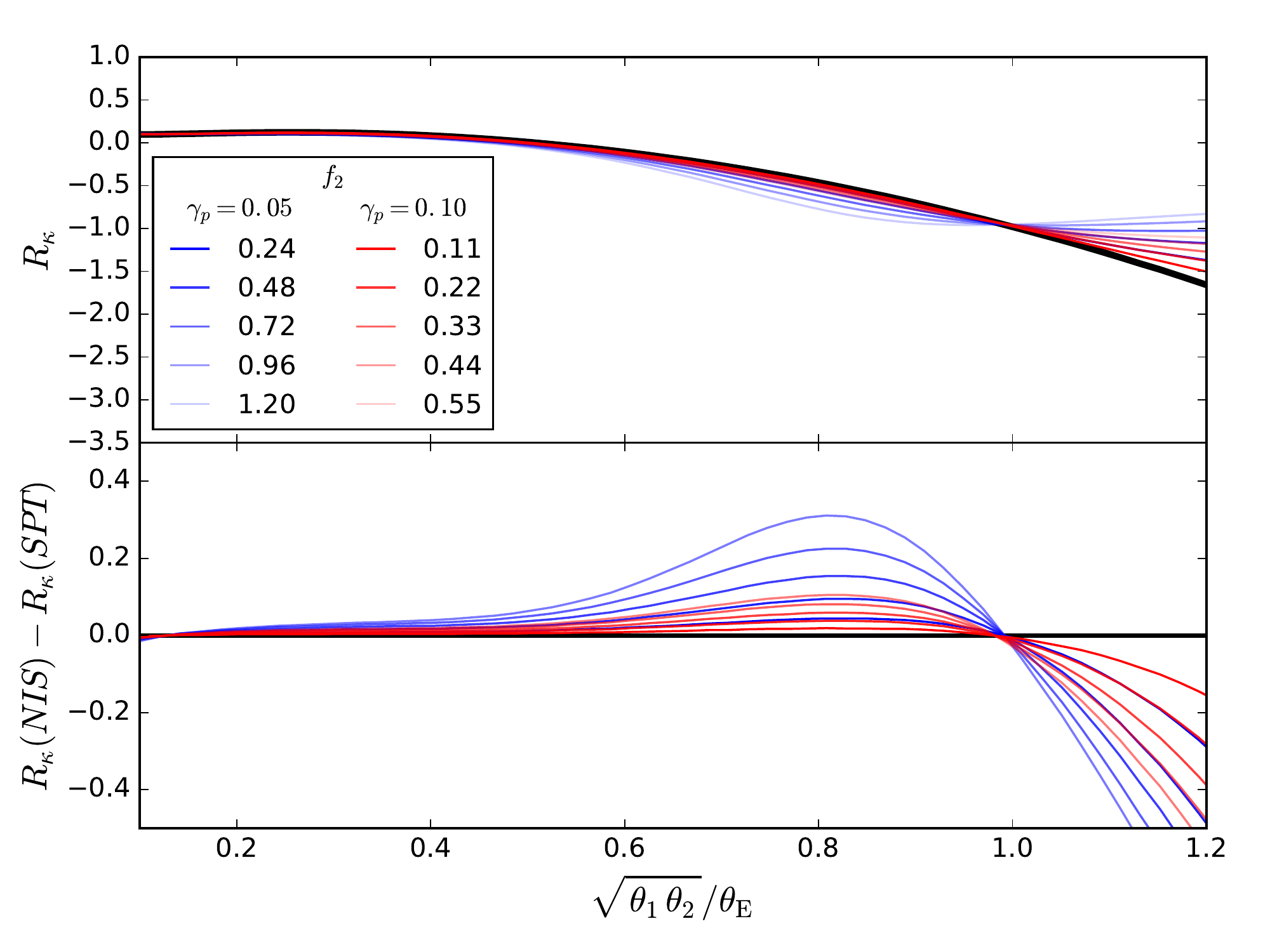}
	\caption{{\bf Top: } Quantity $R_\kappa(\theta)$
          (Eq.~\ref{eq:Rk}) calculated for an NIS with external shear
          \gp\,(cf. Sect.~\ref{sc:Sc4}; solid black) and for various
          SPT-transformed models with SPT of the form
          1+$f_2/2\,(\beta/\theta_{\rm E})^2$
          (Eq.~\ref{eq:deffct}). The range of positive values of
          $f_2$ allowed by $|\Delta\,\avec_{\rm
            {max}}|<5\times10^{-3}\,\theta_{\rm E}$
          (Fig.~\ref{pic:f2gp}) is explored for two different choices
          of the shear: $\gamma_{\rm p}=0.05$ (blue) and
          $\gp=0.1$ (red). While $R_{\kappa}$ is conserved under an MST, it
          is not under an SPT, with deviation that can reach tens of
          percents. {\bf Bottom:} For each curve of the top panel, we
          show the difference between $R_\kappa$ of the original NIS
          model and of the SPT transformed model. }
	\label{pic:RkMST}
\end{figure}

Noting that the MST yields a multiplication of $1-\kappa(\theta)$ by
$\lambda$, the ratio
\begin{equation}
	R_\kappa(\theta)=\frac{1-\ave{\kappa}(\theta)}{\ave{\kappa}'(\theta)}\;,
	\elabel{Rk}
\end{equation}
is well defined for monotonically decreasing mass profiles and
invariant under the MST. Figure~\ref{pic:RkMST} shows $R_\kappa$ for
the NIS and various SPT transformed models. The variations of
$R_\kappa$ are particularly significant above one Einstein radius, in
regions where the SPT-transformed profiles deviate also more strongly
from the original profile. Despite the fact that $\hat{\kappa}$ deviates from
$\kappa_{\rm orginal}$ by more than 20\% within $\tE$ when $f_2 = 1.2$
(Fig.~\ref{pic:kapradial}), the most extreme changes of
$R_\kappa(\theta)$ reach no more than $\sim$ 10\% within one Einstein
radius. As expected, $R_{\kappa}$ deviates more strongly from the
original profile when $|f_2|$ is large. Negative values of $f_2$ (not
shown) are qualitatively similar (but mirrored
w.r.t. $R^{\rm NIS}_\kappa$) to the situation encountered for positive
$f_2$. However, at radii $\theta \sim \,1.2 \, \theta_{\rm E}$,
$\hat{\kappa}$ stops being monotonically decreasing, and $R_\kappa$
diverges.

Another possibility to characterize the radial profile change is
through the aperture mass (see Schneider 1996). Consider a function
$U(\theta;\theta_0) = \theta_0^{-2}\,u(\theta/\theta_0)$ such that
$u(x)$ is non-zero only for $x\le 1$; hence, $\theta_0$ characterizes
the range of support of $U(\theta;\theta_0)$. Furthermore, we require
that the filter function $U$ has a vanishing two-dimensional integral
over its support, which means that
\[
	\int_0^1\d x\; x\;u(x)=0 \; .
\]
Then we define the aperture mass as
\begin{equation}
	M_{\rm ap}(\theta_0)=\int \d^2\theta\; \kappa(\vc\theta)\,
	U(|\vc\theta|;\theta_0)
	=2\pi \int\d\theta\;\theta\,\ave{\kappa}(\theta)\,
	U(\theta;\theta_0) \;.
	\elabel{Map}
\end{equation}
The mass-sheet transformation (\ref{eq:MST}) leads to multiplication
of $M_{\rm ap}$ by a factor $\lambda$, whereas the additive term in
(\ref{eq:MST}) drops out, due to the compensated nature of the filter
function $U$. Thus, if we consider the ratio of the aperture mass for
two different scale lengths $\theta_0$, the factor $\lambda$ drops
out, and this ratio $M_{\rm ap}(\theta_1)/M_{\rm ap}(\theta_2)$ is
invariant under the MST.

Consider again a power-law density profile,
$\ave{\kappa}(\theta)=(1-s/2)(\theta/\theta_{\rm E})^{-s}$, with
$0<s<2$, where $\theta_{\rm E}$ is the Einstein radius in case of
axi-symmetry. Then,
\[
	M_{\rm ap}=(2-s)\pi \rund{\frac{\theta_0}{\theta_{\rm E}}}^{-s}
	\int_0^1\d x\;x^{1-s}\,u(x)\;,
\]
and $M_{\rm ap}(\theta_1)/M_{\rm
  ap}(\theta_2)=(\theta_1/\theta_2)^{-s}$. We thus define the
effective slope
\begin{equation}
	s_{\rm eff}:=\frac{\ln\eck{M_{\rm ap}(\theta_1)/M_{\rm
	  ap}(\theta_2)}}{\ln (\theta_2/ \theta_1)}\;,
	\elabel{seff}
\end{equation}
so that for a mass profile of the form
$\ave{\kappa}(\theta)=a+b\theta^{-s}$, $s_{\rm eff}=s$.

One can think of a number of appropriate weight functions $u(x)$ and
aperture scales $\theta_i$ to characterize the modified mass
profile. The simplest form would be the sum of two delta functions,
$u(x)=\delta(x-x_0)-x_0\delta(x-1)$, with $x_0<1$, for which $M_{\rm
  ap}(\theta_0) = 2\pi
x_0\eck{\ave{\kappa}(x_0\theta_0)-\ave{\kappa}(\theta_0)}$. Furthermore,
choosing $\theta_2=\theta_1/x_0$, the ratio of aperture masses becomes
\begin{equation}
	\frac{M_{\rm ap}(\theta_1)}{M_{\rm ap}(\theta_1/x_0)}
	=\frac{\ave{\kappa}(x_0 \theta_1)-\ave{\kappa}(\theta_1)}{\ave{\kappa}(\theta_1)-\ave{\kappa}(\theta_1/x_0)}\;.
\end{equation}
In the case of $(1-x_0)\ll 1$, the expression (\ref{eq:seff}) becomes 
\begin{equation}
	s_{\rm eff}=-1-\frac{\theta_1 \ave{\kappa}''(\theta_1)}
	  {\ave{\kappa}'(\theta_1)} +{\cal O}([1-x_0]^2)\;.
\end{equation}
Hence, we see that in this case $s_{\rm eff}$ depends just on the
ratio of second to first derivative, and reduces to $s$ for a
power-law mass profile with slope $s$.

More practical choices of $u$ would be such that the profile is probed
over an annulus around the Einstein radius $\theta_{\rm E}$. For
  example, one could use the compensated filter function

\begin{equation}
	u(x) = \begin{cases} \frac{1}{x}(x-x_0) (2x-x_0-1) (x-1) &\mbox{for}\ \ \ x_0\le x\le 1 \\ 0 & \mbox{else}\end{cases} 
\label{eq:ux}
\end{equation}

\begin{figure}[htbp]
	\centering
	\includegraphics[width=.5\textwidth]{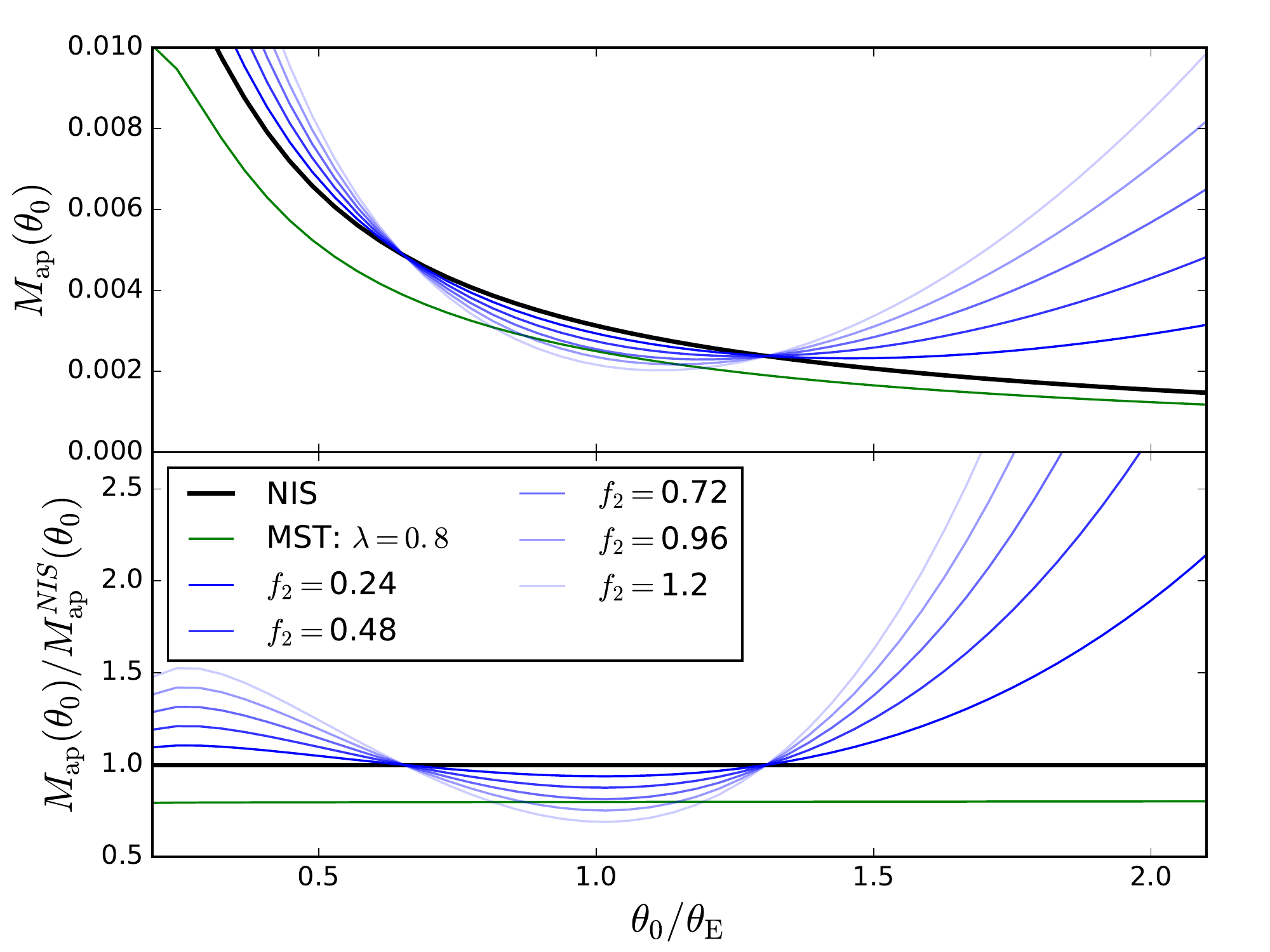}
	\caption{{\bf Top}: $M_{\rm ap}(\theta_0)$ (Eq.~\ref{eq:Map})
          as a function of the ``aperture'' $\theta_0$. The filter
          function $u(x)$ defined by Eq.\ts (\ref{eq:ux}), using $x_0
          = 0.5$, has been used such that for $M_{\rm ap}(\theta_0)$,
          the annulus $[\theta_0/(2\theta_E), \theta_0/\theta_E]$ is
          probed. The black curve shows $M_{\rm ap}$ for the NIS
          profile and the blue curves for the SPT-transformed profiles
          with $\gamma_{\rm p}=0.05$ and various values of $f_2$. The green
          curve shows $M_{\rm ap} (\theta_0)$ for an MST transformed
          version of the NIS profile. {\bf Bottom:} Ratio between
          $M_{\rm ap}$ derived for the various transformed profiles
          and for the original NIS profile.  }
	\label{pic:Map}
\end{figure}

Figure~\ref{pic:Map} shows $M_{\rm
  ap}$ as a function of $\theta_0$, fixing $x_0 = 1/2$ in
(\ref{eq:ux}). As expected, for two profiles transformed into each
other via an MST, the ratio of aperture masses is independent of
$\theta_0$ and equals $\lambda$. Conversely, $M_{\rm ap}(\theta_0)$ of
the SPT-transformed profiles intersects the aperture mass ``function''
of the original profile (i.e. NIS), and the ratio between the two
curves changes with $\theta_0$. The radius at which the curves
intersects is almost independent of the value of $f_2$. This can
easily be deduced from the apparent self-similarity of the
SPT-transformed mass density profiles (Fig.~\ref{pic:kapradial}) for
various values of $f_2$.
	
\begin{figure}[htbp]
	\centering
	\includegraphics[width=.5\textwidth]{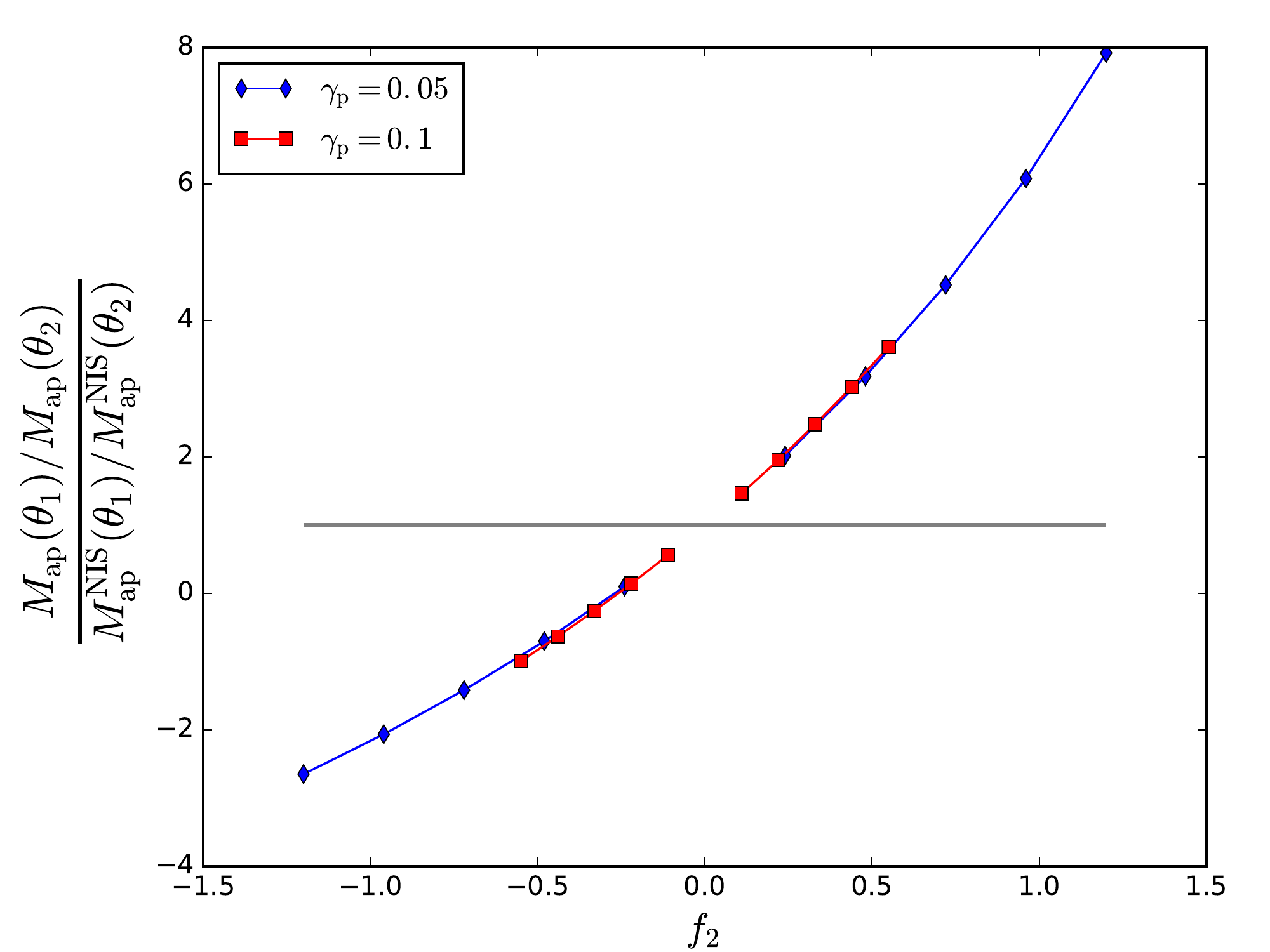}
	\caption{Ratios of aperture mass between $\theta_1 =
          2\,\theta_E$ and $\theta_2 = \theta_E$ for SPT-transformed
          profiles with various values of $f_2$. The ratios of
          aperture masses are normalized by the corresponding aperture
          ratios estimated for the original NIS profile (horizontal
          bar). The blue diamonds are for a shear $\gamma_{\rm p} =
          0.05$, and the red squares when $\gamma_{\rm p} = 0.1$. } 
	\label{pic:Mapratio}
\end{figure}

Figure~\ref{pic:Map} motivates a choice of radii corresponding to
extrema of $M_{\rm ap}(\theta_0)$ to calculate aperture mass ratios,
such as $\theta_1=2\,\theta_{\rm E}$ and $\theta_2=\theta_{\rm
  E}$. Then, $M_{\rm ap}(\theta_1)$ will probe the annulus
$\theta_{\rm E} < \theta < 2\,\theta_{\rm E}$, while for $\theta_2
\sim \theta_{\rm E}$ the annulus $0.5\,\theta_{\rm E}\le
\theta\le\theta_{\rm E}$ would be probed. Figure~\ref{pic:Mapratio}
shows normalized aperture mass ratios $M_{\rm ap}(\theta_1)/M_{\rm
  ap}(\theta_2)$ for the various SPT-transformed profiles studied in
the previous section. We see that larger aperture mass ratios are
found for larger values of $f_2$. In addition, the ratio depends only
weakly of the shear amplitude $\gp$, which means that the radial
deformation of the mass profile produced by the SPT is mostly governed
by the amplitude of $f_2$.

\subsection{A specific lens model}
\begin{figure}[htbp]
	\centering
	\includegraphics[width=.5\textwidth]{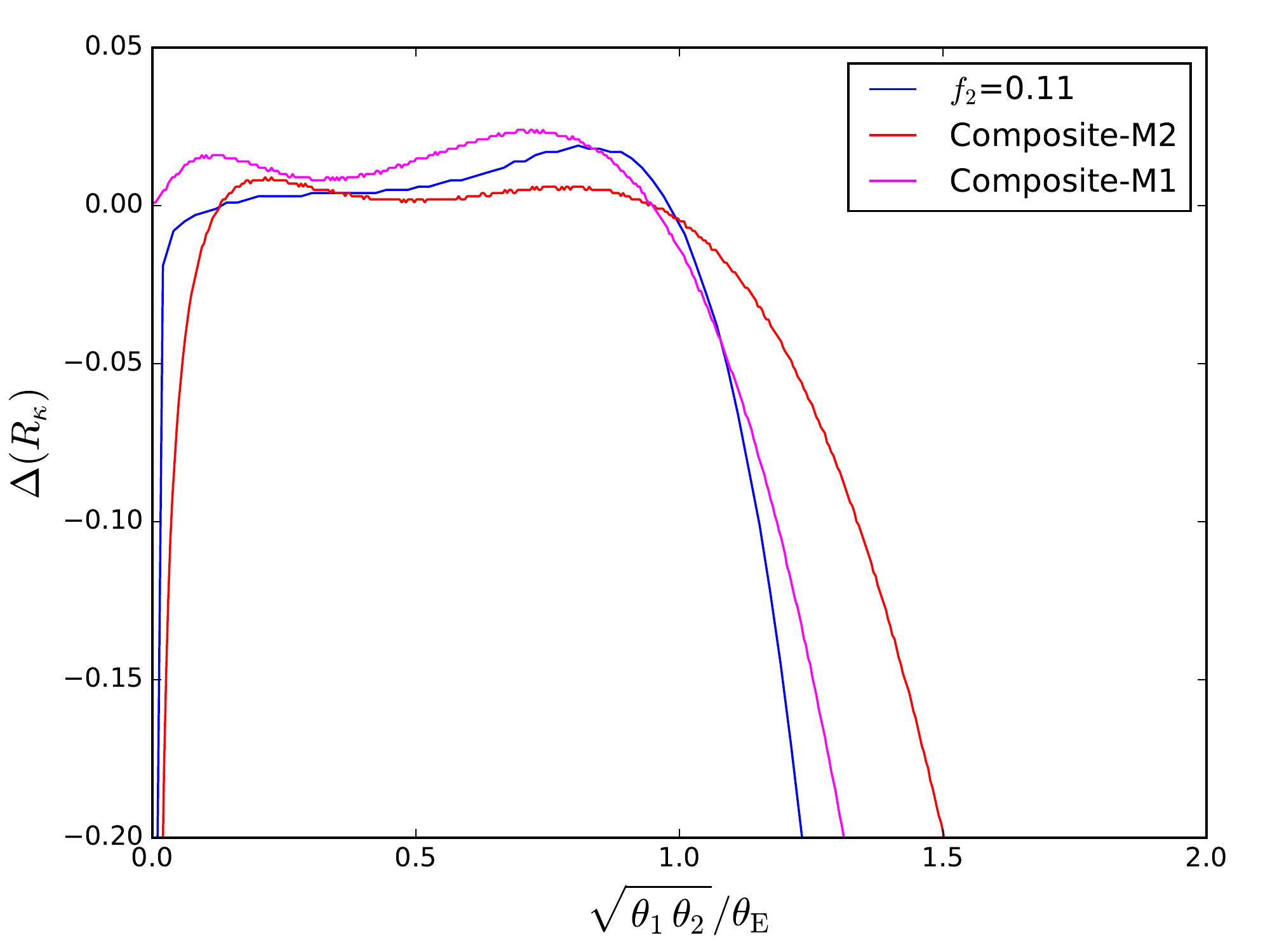}
	\caption{Difference between $R_\kappa$ calculated for three
          different pairs of profiles: In blue, an NIS and an
          SPT-transformed model with $f_2=0.11$ and $\gamma_p=0.1$; in
          magenta, a composite Hernquist+gNFW model and a power-law
          model (M1); in red, Hernquist+gNFW model and a cored
          power-law model (M2). The shape of $\Delta R_\kappa$ for the
          models presented in SS13 and SS14 are qualitatively similar
          to that observed for the fiducial SPT model presented in
          Sect.~\ref{sc:Sc4}.  }
	\label{pic:RkSS13}
\end{figure}

Here, we apply the previous tests to three mass density profiles
studied in SS13
and SS14, and used in those papers to illustrate
degeneracies produced by the MST and the SPT. The reference model is
a composite model constituted of the sum of a (spherically symmetric)
Hernquist component to describe the baryonic component, and a
(spherical) generalized Navarro-Frank-White (gNFW) density profile to
describe the dark matter component of the galaxy. In addition, an
external shear of amplitude $\gamma_{\rm p} = 0.1$ is considered. Complex sets
of lensed images from an ensemble of sources were generated with that
model, and found to be all  
equally well reproduced by two (single) power-law profiles: a global
power law, with an almost isothermal density slope
$\gamma^{\prime}=1.98$ (hereafter model M1), and a local power-law
profile with a core radius $\theta_c = 0.1^{\prime\prime}$ and a slope
$\gamma^{\prime}=$ 2.2 (hereafter model M2). We have applied the tests
introduced in the previous subsection to these profiles to identify
the nature of the degeneracy between the
models. Figure~\ref{pic:RkSS13} shows the difference $R_\kappa$
between the original profile and transformed ones, i.e., $\Delta
R_\kappa = R_\kappa({\rm original})-R_\kappa(\rm {transformed})$. For
comparison, we also show $\Delta R_\kappa$ obtained for an SPT with
$f_2=0.11$. This figure suggests that indeed, this degeneracy is
similar to an SPT.

\begin{figure}[htbp]
	\centering
	\includegraphics[width=.5\textwidth]{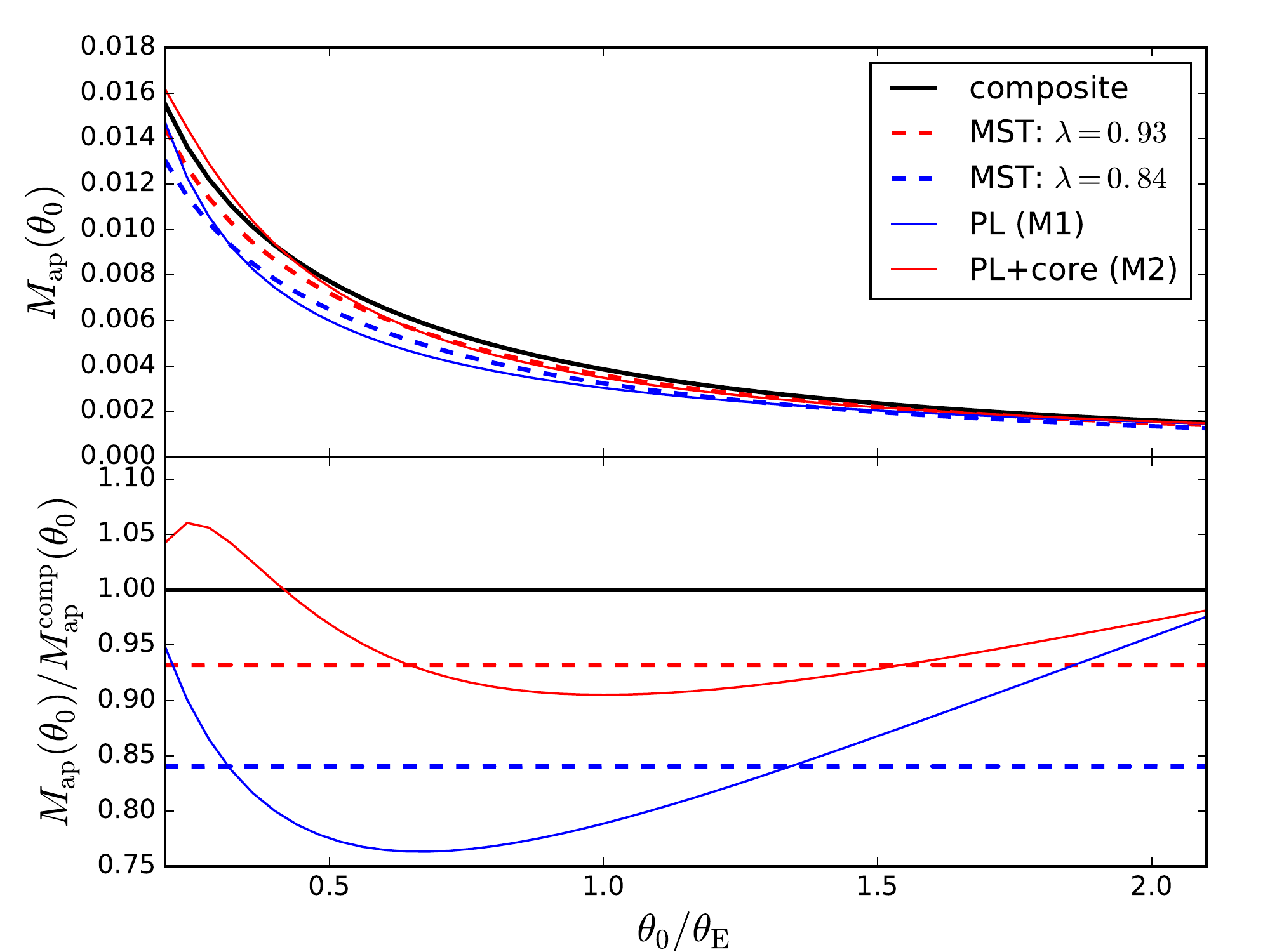}
	\caption{{\bf Top:} $M_{\rm ap}(\theta_0)$ (Eq.~\ref{eq:Map})
          as a function of $\theta_0$, for the composite
          Hernquist+gNFW model (black), for the power law model M1
          (blue), and the cored power-law M2 (red). Dashed red (blue)
          profile shows $M_{\rm ap}(\theta_0)$ for an MST transformed
          version of the composite model with $\lambda=0.93$
          (resp. 0.84) {\bf Bottom:} Ratio of $M_{\rm ap}(\theta_0)$
          between the ``transformed'' models and the composite. The
          dashed curves correspond to MST-transformed versions of the
          composite model, and represent the contribution of the MST
          to M1 and M2. The solid red and blue curves suggest that the
          remaining of the degeneracy can be associated with an SPT. }
	\label{pic:MapSS}
\end{figure}

The other diagnostic we present consists in calculating the aperture
mass $M_{\rm ap}$ of the profiles. Figure~\ref{pic:MapSS} shows
$M_{\rm ap}$ as a function of $\theta_0$. In addition to the aperture
mass calculated for the individual profiles, we also show the aperture
mass corresponding to two different MST-transformed mass density
profiles. As explained in SS13, model M1 is close to an MST transformed
version of the composite model\footnote{In SS13, we reported
  $\lambda=\beta_{\rm fid}/\beta_{\rm PL}=1.19$, which is the inverse
  of $\lambda$ used here that is such that $\kappa_{\rm PL} =
  \lambda\,\kappa_{\rm fid} + (1-\lambda)$.} with $\lambda=0.84$. On
the other hand, Fig. 4 of SS14 shows that the MST contribution to M2
corresponds to $\lambda=0.932$. Figure~\ref{pic:MapSS} is
qualitatively similar to Fig.~\ref{pic:Map} but there is an offset of
$M_{\rm ap}$ for M1 and M2 compared to the composite model. The
reason is probably that the M1 and M2 profiles are transformed versions of
the composite model via both an MST {\it and} an SPT. The MST
contribution with $\lambda=0.84$ is larger for M1 than for M2, for
which $\lambda\sim0.93$.


\section{\llabel{Sc6}Discussion -- Implications of the SPT for strong lensing}
%
%
%
In this paper we have studied several aspects of the SPT, an
invariance transformation of the deflection angle that leaves all
multiple images properties of gravitational lenses invariant. The
central question, of whether there exists a gravitational lensing
potential which gives rise to a deflection angle sufficiently close to
the SPT-transformed one (which in general is not curl free) has been
explored for a particular class of lens models, namely an NIS with
external shear and an SPT given as a radial stretching of the source
plane. The radial stretching deformation was chosen such that the
classical MST did not contribute in altering the original deflection
since we are only interested in higher-order effects that go beyond
the well known MST. This example has shown that, for a large range of
parameters pairs (external shear and distortion parameter of the
radial stretching) there indeed exist lensing potentials whose
associated deflection is sufficiently close to the one obtained from
the SPT that these two cannot be distinguish observationally. We
conducted this study by formulating an action as the integral over the
squared difference of these two deflection angles, yield a von Neumann
problem. We gave a detailed description of how this problem can be
solved; these methods are expected to be useful for future theoretical
studies and applications of the SPT.

We have considered only one criterion for the validity of an
SPT, namely that the corresponding gradient deflection field does not
deviate from the SPT-transformed deflection by more than $5\times
10^{-3}\tE$. Changing this observationally motivated limit to a
different value will modify the space of allowed parameter
combinations. For the example considered here, we expect that the
allowed range of the stretching parameter $f_2$ for a given external
shear will be proportional to the allowed maximum deviation of these
two deflection angles.

The properties of the mass distribution resulting from an SPT were
also studied in detail. In contrast to the MST, which is a special
case of an SPT, the more general SPT gives rise to non-monotonic
changes in the radial mass profile, and to the generation of a finite
ellipticity even if the original mass distribution was
axi-symmetric. Hence, the SPT offers a much larger range of mass
profile modifications which leave all strong lensing observables
invariant, than does the MST. This more complex class of invariance
transformation is of particular interest because it may be of great
relevance when trying to fit real lens system (which are expected to
have a rather complex mass profile; see, e.g., \citet{Dandan16} with simple lens models. The fact that simple mass models
yield satisfactory fits even in cases with a rich observed image
structure may be a consequence of some SPT which transforms the
deflection of the true mass distribution into that of a simple mass
model.  Ignoring the potential complexity of the real mass
distribution, and thus the possibility that the SPT may be acting, may
lead to biases in estimates of physical parameters of the lens system.

We have defined several diagnostic quantities which can distinguish a
general SPT from a pure MST. Applying these diagnostics to the special
case of nearly degenerate lens models studies in two earlier papers, we
conclude that this degeneracy can to a large degree be accounted by an
MST, but that a non-negligible contribution comes from a more general
SPT. Hence, an SPT has been found `empirically', even before the
concept of the SPT was developed. In that sense, the SPT is not just a
`theoretical possibility' for obtaining different but observationally
equivalent mass models, but describes degeneracies which actually
occur in real lens modeling.

%
%
\begin{appendix}
\section{\llabel{ApA}Practical integration of Eq.\ts
  (\ref{eq:Hatilde}) in a circular region}
Calculating the deflection angle $\nabla \tilde\psi$ in Eq.\ts
(\ref{eq:vNSol}) by integrating the product $\nabla_\vt H\, \hat\kappa$
over the circle poses a challenge, due to the pole of the first term
in Eq.\ts (\ref{eq:nabH}). To integrate over this pole, polar coordinates
centered on the pole position $\vc\vt$ need to be chosen. This can be
done by a translation of the integration variable to $\vc
x=\vc\theta-\vc\vt$, and integrating in the polar coordinates of $\vc
x$. However, the integration range of the polar angle will depend on
$|\vc x|$, according to the geometrical overlap of circles centered on
the origin and those centered on $\vc\theta$.

A better method is obtained by a conformal mapping of the form
\begin{equation}
	x=\frac{\theta-\vt}{R-\vt^* \theta/R}\;,
	\elabel{contrans}
\end{equation}
where we now use complex notation, i.e., $x$, $\vt$ and $\theta$ are
complex numbers with components $\vt=\vt_1+{\rm i}\vt_2$ etc. and an
asterisk denotes complex conjugation. This conformal mapping maps the
circle $|\theta|<R$ onto the unit circle $|x|<1$, and the singularity
point $\theta=\vt$ is mapped onto the origin $x=0$. For example,
setting $\theta=R{\rm e}^{{\rm i}\vp}$, we get
\[
	x=\frac{R{\rm e}^{{\rm i}\vp}-\vt}{R-\vt^* {\rm e}^{{\rm i}\vp}}
	={\rm e}^{{\rm i}\vp}\frac{R-\vt {\rm e}^{{-\rm i}\vp}}{\rund{R-\vt {\rm e}^{{-\rm i}\vp }}^*}\;,
\]
from which it is immediately seen that $|x|=1$. The inverse of the transformation (\ref{eq:contrans}) is readily obtained,
\begin{equation}
	\theta=\frac{R x+\vt}{1+\vt^* x/R}\;,
	\elabel{contransinv}
\end{equation}
from which one can easily check that the unit circle $|x|=1$ is mapped
onto the circle $|\theta|=R$. In components, Eq.\ts
(\ref{eq:contransinv}) reads 
\begin{eqnarray}
	\theta_1&=&\frac{R
	  x_1+\vt_1(1+|x|^2)+[x_1(\vt_1^2-\vt_2^2)+2\vt_1\vt_2x_2]/R}
	{1+2\vc\vt\cdot\vc x/R+|\vc\vt|^2 |\vc x|^2/R^2}\;,
	\nonumber \\
	\theta_2&=&\frac{R
	  x_2+\vt_2(1+|x|^2)+[2\vt_1\vt_2 x_1 - x_2(\vt_1^2-\vt_2^2)]/R}
	{1+2\vc\vt\cdot\vc x/R+|\vc\vt|^2 |\vc x|^2/R^2}\;.
\end{eqnarray}
The Jacobi determinant of the transformation $x \to \theta$, needed for the integration, is
\begin{equation}
	\abs{ \frac{\partial \vc\theta}{\partial \vc x}}
	=\frac{R^2 (R^2-|\vc\vt|^2)^2}{(R^2+2 R \vc\vt\cdot\vc x+|\vc\vt|^2
	  |\vc x|^2)^2} \;,
\end{equation}
which is non-zero for all $\vc x$ inside the unit circle and $|\vc\vt|<R$. As a sanity check, we calculate the area of the circle in the transformed coordinates,
\begin{eqnarray}
	R^2\pi&=&\int_{\cal U}\d^2\theta = \int_{\cal C}\d^2 x\;\abs{
	  \frac{\partial \vc\theta}{\partial \vc x}}  \\
	&=&R^2 (R^2-|\vc\vt|^2)^2\int_0^1\d x\;x
	\int_0^{2\pi}\frac{\d \vp}{(R^2+2 R \vc\vt\cdot\vc x+|\vc\vt|^2
	  |\vc x|^2)^2}\;.\nonumber
\end{eqnarray}
The inner integral yields $2\pi (R^2+|\vc x|^2|\vc\vt|^2)/(R^2-|\vc x|^2|\vc\vt|^2)^3$, the outer integral then gives $\pi/(R^2-|\vc\vt|^2)^2$, and we re-obtain the area $\pi R^2$.

In complex notation, the singular term in Eq.\ts (\ref{eq:nabH}) reads
\[
	\frac{1}{(\vt-\theta)^*}=\frac{x + |x|^2\vt/R}{R |x|^2
	  \rund{|\vt|^2/R^2-1}}\;,
\]
yielding
\begin{equation}
	\abs{\frac{\partial \vc\theta}{\partial \vc x}}
	\frac{\vc\vt-\vc\theta}{|\vc\vt-\vc\theta |^2}
	=\frac{R^2(|\vc\vt|^2-R^2)}{\rund{R^2+2 R \vc\vt\cdot\vc x+|\vc\vt|^2
	  |\vc x|^2}^2}\rund{\frac{R}{\abs{\vc x}^2}\vc x +\vc\vt}\;.
\end{equation}
We can check the consistency of this expression by calculating the deflection angle of a uniform disk with surface mass density $\kappa_0$, which reads
\begin{eqnarray}
	\vc\alpha(\vc\vt)&=&\frac{\kappa_0}{\pi}\int_{\cal U}\d^2\theta\;
	\frac{\vc\vt-\vc\theta}{|\vc\vt-\vc\theta |^2}
	=\frac{\kappa_0}{\pi}\int_{\cal C}\d^2 x\;
	\abs{\frac{\partial \vc\theta}{\partial \vc x}}
	\frac{\vc\vt-\vc\theta}{|\vc\vt-\vc\theta |^2} \nonumber \\
	&=&\frac{R^2\rund{|\vc\vt|^2-R^2} \kappa_0}{\pi}\int_0^1\d
	x\;x
	 \nonumber\\
	&\times&
	\int_0^{2\pi} \frac{\d\vp}{\rund{R^2+2 R \vc\vt\cdot\vc x+|\vc\vt|^2
	  x^2}^2}\rund{\frac{R}{x^2}\vc x +\vc\vt} \\
	&=&\frac{R^2\rund{|\vc\vt|^2-R^2} \kappa_0}{\pi}\int_0^1\d
	x\;x\, \rund{\frac{-2 \pi}{\rund{R^2-|\vc\vt|^2 x^2}^2}}\vc\vt  \nonumber\\
	&=& \frac{R^2\rund{|\vc\vt|^2-R^2} \kappa_0}{\pi} 
	\rund{\frac{\pi}{R^2 \rund{|\vc\vt|^2-R^2}}}\vc\vt=\kappa_0\,\vc\vt\;, \nonumber
\end{eqnarray}
as expected.
\section{Proof of the relation (\ref{eq:A5}) for a circular region}
In this section we use again the complex notation for two-dimensional vectors, in terms of which the vector field (\ref{eq:nabH}) reads
\begin{equation}
	\nabla_\vt H(\vc\vt;\vc\theta) \to
	\frac{1}{2\pi}\rund{\frac{1}{\vt^*-\theta^*}-\frac{\theta}
	    {(R^2-\vt^*\theta)}-\frac{\vt}{R^2}}\;.
\end{equation}
Therefore, we obtain for the fields $\vc A$ and $\vc B_{\rm in}$ the complex expressions
\begin{eqnarray}
	A(\vt)&=&-\frac{1}{\pi}\int_{\cal U} \!\! \d^2\theta\;\kappa(\vc\theta)\,
	\rund{\frac{\theta}{(R^2-\vt^*\theta)}+\frac{\vt}{R^2}}\;, \nonumber \\
	B_{\rm in}(\vt)&=&\frac{1}{2\pi}\int_{\partial \cal U}\!\!\!\d
	s\;\rund{\frac{1}{\vt^*-\theta^*}-\frac{\theta}
	    {(R^2-\vt^*\theta)}-\frac{\vt}{R^2}}\, a_{\rm in}\;,
	\elabel{Ap2}
\end{eqnarray}
where $a_{\rm in}=\vc\alpha_{\rm in}\cdot n$, defined on the boundary
of $\cal U$. We first calculate this product, noting that for a
boundary point $R\,{\rm e}^{{\rm i}\vp}$ of the circle, $n={\rm
  e}^{{\rm i}\vp}$. Specializing the complex expression for
$\vc\alpha_{\rm in}$ as given in Eq.\ts (\ref{eq:A1}) to a point on
the boundary, $\vt=R n$, we obtain 
\begin{eqnarray}
	a_{\rm in}(R n)&=&\vc\alpha_{\rm in}\cdot n
	=\frac{\alpha_{\rm in}/n+\alpha_{\rm in}^* n}{2}\nonumber\\
	&=&\frac{1}{2\pi}\int_{\cal U} \!\! \d^2\theta\;\kappa(\vc\theta)
	\rund{\frac{1}{R-n \theta^*}+\frac{n}{R n-\theta}}\;.
	\elabel{Ap3}
\end{eqnarray}
We now insert this expression into Eq.\ts (\ref{eq:Ap2}). The integral over
the boundary is written as $\d s=R\, \d\vp=-{\rm i}\,R\, \d n/n$, and the
integral over $n$ extends over the unit circle. With $\theta=R n$, we
then find 
\begin{eqnarray}
	B_{\rm in}(\vt)&=&\frac{R}{2\pi}\int_{\cal U} \!\!
	\d^2\theta\;\kappa(\vc\theta) \,\frac{-{\rm i}}{2\pi}
	\oint\frac{\d n}{n}\rund{\frac{1}{R-n\theta^*}+\frac{n}{R
	    n-\theta}}\nonumber \\
	&&\times
	\rund{\frac{n}{n\vt^*-R}-\frac{n}{R-n\vt^*}-\frac{\vt}{R^2}}\;.
	\elabel{A4.2}
\end{eqnarray}
The inner integrand is an analytic function of $n$ inside the unit circle, except at the poles at $n=0$ and at $n=\theta/R$ (note that $\vt$ and $\theta$ are both inside the circle). Applying the theorem of residue, the integral can thus be evaluated. The first pole yields a contribution $-\vt/R^3$, whereas the second pole results in the expression 
\[
	\frac{\theta}{(\theta\vt^*-R^2)R} -
	\frac{\theta}{(R^2-\theta\vt^*)R}-\frac{\vt}{R^3}\;.
\]
Adding up these two contributions then yields
\begin{equation}
	B_{\rm in}(\vt)=\frac{1}{\pi}\int_{\cal U} \!\!
	\d^2\theta\;\kappa(\vc\theta)\rund{-\frac{\theta}
	    {(R^2-\vt^*\theta)}-\frac{\vt}{R^2}} \;,
\end{equation}
which we see agrees with the expression for $A$ in
Eq.\ts (\ref{eq:Ap2}). Thus we have shown explicitly that for the kernel
function (\ref{eq:Hcirc}), the relation (\ref{eq:A5}) holds.
\section{Conservation of mass inside the Einstein radius under SPT}
\label{sec:masscons}
Starting from (\ref{eq:khatSS14}) we can infer the mass inside the
Einstein radius by performing an integration up to \tE. First, we
consider the case of an axisymmetric lens model, i.e., \(\gp =
0\). Thus, the integral we have to solve is 
\begin{align}
	\hat{M}_{\gp=0}(\le\tE) &= 2 \int_0^{\tE} \d \theta \; \theta \, \khat(\theta) \nonumber\\
	= M(\le\tE) &+ f_2 \int_0^{\tE} \d \theta \; \theta^3 ( 1 - \kbar )^2 [ 3 (\kappa - \kbar) - 2 (1 - \kbar) ] \;.
	\label{eq:MintE} 
\end{align}
To show that mass is conserved in case of an SPT, the integral in
Eq.\ts (\ref{eq:MintE}) has to vanish for arbitray mass profiles
\(\kappa(\theta)\). We note that \kbar \ is given by
\begin{align}
	\kbar (\theta) = \frac{2}{\theta^2} \int_0^\theta \d \theta' \; \theta' \, \kappa(\theta') \;,
\end{align}
from which follows
\begin{align}
	\kbar' &=  \frac{2}{\theta} (\kappa - \kbar) \;,
	\elabel{kbarprime}
\end{align}
of which we will make use in the next step. We perform an integration
by parts for the last term in Eq.\ts (\ref{eq:MintE}),
\begin{align}
	2 \int_0^{\tE} \d \theta \; \theta^3 (1 - \kbar)^3 &= \left[ 2 \frac{\theta^4}{4} (1-\kbar)^3 \right]^{\tE}_0 + 2 \int_0^{\tE} \d \theta \; \frac{\theta^4}{4} \, 3 (1-\kbar)^2 \kbar' \nonumber\\
	&= 3 \int_0^{\tE} \d \theta \; \theta^3 \, (1-\kbar)^2 (\kappa - \kbar) \;,
\end{align}
where we used \( \kbar (\tE) = 1 \). This result matches exactly the
first term in the integral of Eq.\ts (\ref{eq:MintE}). Hence, in the
case \(\gp = 0\) the mass inside the Einstein ring is conserved under
an SPT.

Next, we consider the case for \(\gp \ne 0\). To integrate \khat \
over the Einstein radius we make use of the result above, i.e.,
\mbox{\(\hat{M}_{\gp=0}(\le\tE) = M ( \le \tE )\)}. Since the two
latter parts of Eq.\ts (\ref{eq:khatSS14}) are proportional to \( \cos
(2\phi) \) or \( \cos (4\phi) \), respectively, they do not contribute
to an integral over a circular area. Hence, we calculate
\begin{align}
	\hat{M}(\le\tE) &= 2 \int_0^{\tE} \d \theta \; \theta \khat(\theta) \nonumber\\
	= M(\le\tE) &+ f_2 \gp^2 \int_0^{\tE} \d \theta \; \theta^3 [ 2 (\kappa - \kbar) - 4 (1 - \kbar) ] \;.
	\label{eq:MintEfull}
\end{align}
Again, we consider first the last term in the integral of
Eq.\ts (\ref{eq:MintEfull}), and we make use of Eq.\ts
(\ref{eq:kbarprime}) and \( \kbar 
(\tE) = 1 \). Then, we obtain
\begin{align}
	4 \int_0^{\tE} \d \theta \; \theta^3 (1 - \kbar) &= \left[ 4 \frac{\theta^4}{4} (1-\kbar) \right]^{\tE}_0 + 4 \int_0^{\tE} \d \theta \; \frac{\theta^4}{4} \, \kbar' \nonumber\\
	&= 2 \int_0^{\tE} \d \theta \; \theta^3 \, (\kappa - \kbar) \;. 
\end{align}
This matches the first term in the integral exactly and therefore, the
integral vanishes. Thus, also in the case of an external shear \gp \
the mass enclosed in the Einstein ring is conserved, i.e., \mbox{\(
  \hat{M}(\le\tE) = M(\le\tE) \)}, independent of the choice of mass
\mbox{model \(\kappa\)}.
%
%
\end{appendix}
\begin{acknowledgements}
	Part of this work was supported by the German \emph{Deut\-sche For\-schungs\-ge\-mein\-schaft, DFG\/} project number SL172/1-1.\\
	Sandra Unruh is a member of the International Max Planck Research School (IMPRS) for Astronomy and Astrophysics at the Universities of Bonn and Cologne. Dominique Sluse is supported by a {\it {Back to Belgium}} grant from the Belgian Federal Science Policy (BELSPO).
\end{acknowledgements}
\bibliographystyle{aa}
\bibliography{MSDbib}
\end{document}